\documentclass[letterpaper,twocolumn,10pt]{article}
\usepackage{usenix2019_v3}

\usepackage[utf8]{inputenc}
\usepackage[T1]{fontenc}
\usepackage[australian]{babel}
\usepackage{color,xcolor}
\usepackage{listings}
\usepackage{courier}

\usepackage{amsmath}
\usepackage{marginnote}
\usepackage{colortbl}
\usepackage{multirow}
\usepackage[disable]{todonotes}

\PassOptionsToPackage{hyphens}{url}
\usepackage{balance}
\usepackage{datetime}
\usepackage{enumitem}
\usepackage{latexsym}
\usepackage{subfigure}
\usepackage{algorithm}
\usepackage{algorithmicx}
\usepackage{algpseudocode}
\usepackage{ioa_code_small}
\usepackage{macros}

\usepackage{epstopdf}
\usepackage{booktabs}
\usepackage{multicol}
\usepackage{hyperref}
\hypersetup{
    colorlinks = true,
    linkcolor = {blue},
    linkbordercolor = {blue},
    anchorcolor = {blue},
    citecolor = {blue},
}

\graphicspath{ {images/} }

\newcommand{\parinya}[1]{}

\makeatletter
\newcommand\footnoteref[1]{\protected@xdef\@thefnmark{\ref{#1}}\@footnotemark}
\makeatother

\begin{document}
	
\title{\Large \bf The Attack of the Clones Against Proof-of-Authority}

\author{
Parinya Ekparinya\\University of Sydney 
\and Vincent Gramoli\\University of Sydney 
 \and Guillaume Jourjon\\ Data61-CSIRO 
}



\maketitle

\begin{abstract}
The vulnerability of traditional blockchains have been demonstrated at multiple occasions. 
Various companies are now moving towards Proof-of-Authority (PoA) blockchains with more conventional Byzantine fault tolerance, where a known set of $n$ permissioned sealers among which no more than $t$ are Byzantine seal blocks that include user transactions. Despite their wide adoption, these protocols were not proved correct.

In this paper, we present the Cloning Attack against the two mostly deployed PoA implementations of Ethereum, namely Aura and Clique. The Cloning Attack consists in one sealer cloning its key-value pair into two distinct Ethereum instances that communicate with distinct groups of sealers.
To identify their vulnerabilities, we first specified the corresponding algorithms. We then infer the topology of the largest PoA network, POA Core, through active measurement. We  deploy one testnet for each protocol and demonstrate the success of the attack with only one byzantine sealer. Finally, we propose counter-measures that prevent an adversary from double spending.
\end{abstract}

\section{Introduction}


Ethereum is one of the most popular blockchain systems thanks to the large ecosystem of distributed applications that it executes. Unfortunately, the default Ethereum protocol based on \emph{proof-of-work (PoW)} can fork as it allows distinct blocks to be appended at the same index of the chain. This forking situation can lead to security vulnerabilities, like double spending, if it is not detected early enough~\cite{HKZG15,NKMS16,NG17}. 
Alternative protocols, called \emph{proof-of-authority (PoA)} protocols, that aim at avoiding forks have recently been integrated in the most widely deployed versions of Ethereum, $\lit{parity}$ and $\lit{geth}$, and are currently used world-wide.
PoA has become rapidly popular and is now distributed through major Software-as-a-Service providers 
and used in several blockchain networks~\cite{jpmorganchase_quorum,WLT19,BAF19}. 
Yet, to our knowledge, the level of security offered by PoA protocols has not been properly assessed. 

These PoA consensus protocols, called Aura and Clique, are said to  
use a proof-of-authority
because they restrict the creation of a block to a fixed set of $n$ authority nodes, called \emph{sealers}, among which a maximum of $t < \frac{n}{2}$ can misbehave or be \emph{Byzantine}.
They aim at solving the well-known Byzantine consensus problem~\cite{PSL80}, where among $n$ nodes the \emph{honest} ones agree on a unique block despite the presence of $t<\frac{n}{2}$ Byzantine nodes. 
%
PoA 
gives the sealers the authority to \emph{seal} a block, which consists of signing cryptographically the block. This set of sealers can possibly change over time if a subset of the participants allow it, hence being well suited for dynamic consortia of participants.
%
PoA is an appealing 
solution for critical industries with security 
requirements. 


\begin{figure}
\center{\includegraphics[scale=0.13]{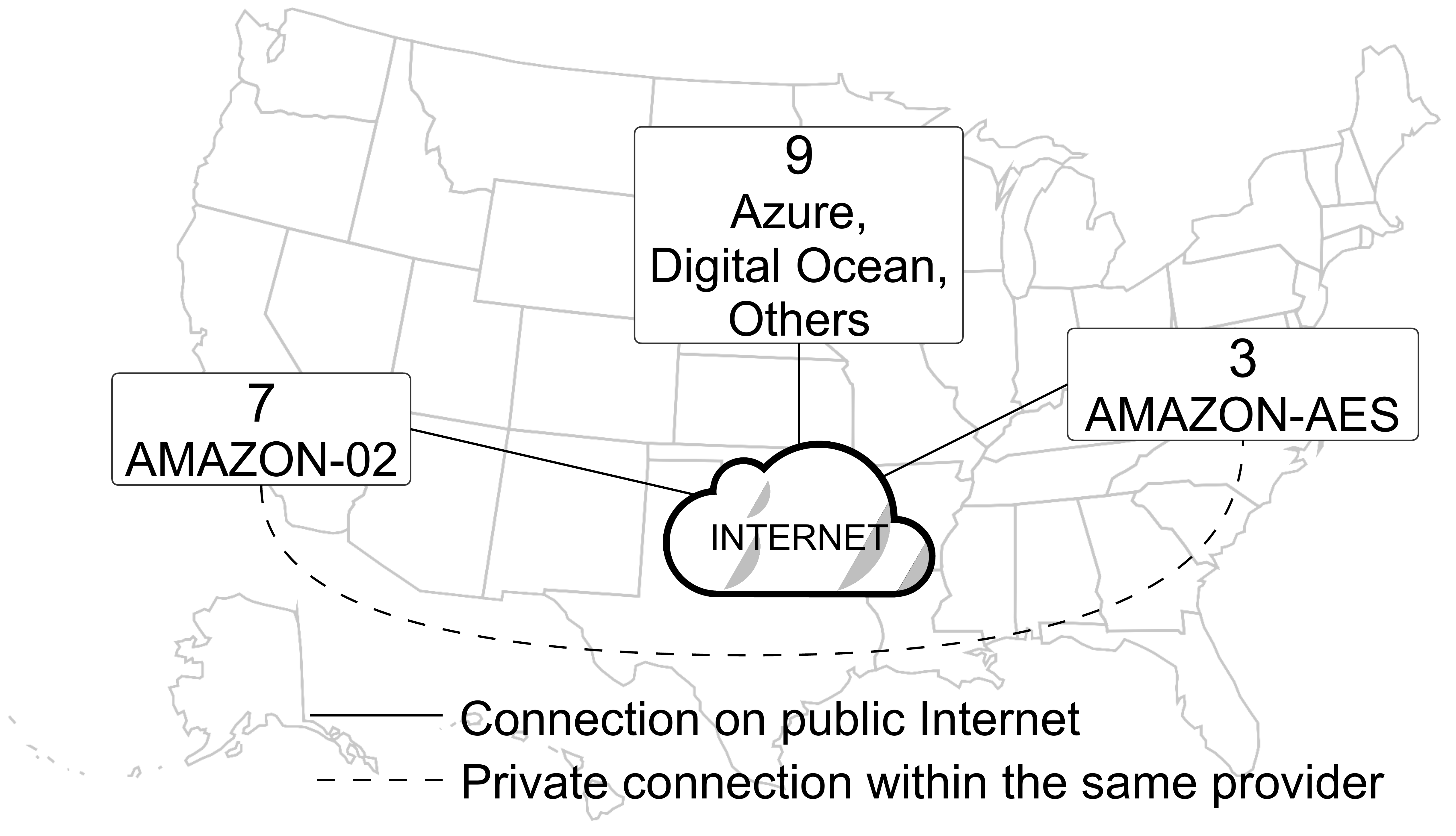}
\caption{Topology of sealers in the POA Core network\label{fig:core}}}
\vspace{-1em}
\end{figure}

For these reasons, PoA recently gained rapid momentum in critical applications~\cite{WLT19,BAF19}.  Industry, such as Lavaa, propose a tracking service to prevent fraud counterfeiting on top of  Aura~\cite{WLT19}. 
Microsoft describes how to deploy Aura ``in production''~\cite{noauthor_ethereum_nodate}.
Amazon Web Services offers PoA through the Clique protocol built in $\lit{geth}$ to its customers~\cite{noauthor_launch_2018}.
They implemented a service that aims at maintaining data privacy and integrity in a multi-tenant scenario. 
%
%
Every day, Internet users exchange digital assets through multiple instances of these two protocols. 
Huawei uses the Apla blockchain platforms based on PoA to develop smart transportation by coupling IoT with blockchain in supply chains and logistics~\cite{noauthor_blockchain_2019}.
Rinkeby is a network of 65 participants offering the Clique service across four continents to its users~\cite{noauthor_rinkeby:_nodate}.
The xDai DPOS network uses an Ethereum 1.0 sidechain based on the Aura consensus protocol to transfer assets~\cite{BAF19}.
Sokol and Kovan are other Ethereum testnets running the Aura protocol~\cite{noauthor_kovan_nodate}.

In particular, Sokol is used to test features before launching them on one of the largest in-production networks called POA Core. 
Figure~\ref{fig:core} illustrates the sealers topology of POA Core that we inferred through active measurements and sealer fingerprinting (cf. Section~\ref{sec:poacore} for details).
%
An interesting theoretical work by De Angelis et al.~\cite{AAB18} indicated that the consistency of Aura can be limited, for example if the clocks are far apart, and that Clique is eventually consistent. 
%
To the best of our knowledge, these algorithms have not been formalized and it is unclear whether an attacker could violate data integrity.

In this paper, we show that, under its required conditions, PoA is not secure even under its required conditions: when a sufficiently large set $V$ of sealers among $n$ of them must seal a block despite a minority $t$ of them being malicious. 
To this end, we design, implement and experiment an attack, called the Cloning Attack, 
against both Ethereum's Aura and Clique consensus protocols that allows us to steal digital assets. 
Our findings inspired by the theory of Byzantine fault tolerance defines precisely the necessary and sufficient conditions $\frac{n + t}{2} < |V| <  n-t$ under which PoA can be safe and live.

The Cloning Attack  
consists of a sealer attacker cloning a private key to convince half of the honest sealers that a transaction is correctly committed before erasing this transaction to double spend its coins.
Thanks to the cloning, to convince half of the honest sealers that transactions are committed, the attacker simply needs to delay messages between two halves of honest sealers.  
Note that this is achieved without knowing the precise locations of sealers but simply through OS and network fingerprinting that allows to identify the autonomous systems the sealers belong to.
In the POA Core example depicted in Figure~\ref{fig:core}, it is sufficient for an Amazon sealer to clone itself on Azure and leverage a 30 second message delay between AWS and the rest of the network to commit conflicting transactions and double spend. \\
%

\noindent\fbox{\begin{minipage}{\dimexpr0.5\textwidth-4.5\fboxsep-2\fboxrule\relax}
{\bf Responsible disclosure:} For ethical reasons, we previously communicated the vulnerability we present here to \emph{(i)}~the security team of Parity Technology, \emph{(ii)}~the Ethereum bug bounty team at 
$\lit{bounty@ethereum.org}$
and \emph{(iii)}~during a non-published presentation at an Ethereum development conference (not revealed here for the sake of anonymity). As a result, both security teams acknowledged the possibility of the attack and the xDai blockchain of the {\sc Posdao} project 
is currently implementing one of the counter-measures of Section~\ref{sec:countermeasure} at \url{https://github.com/poanetwork/parity-ethereum/pull/109} as acknowledged in their white paper~\cite{BAF19}.
\end{minipage}}
\vspace{0.3em}

%
We demonstrate the effectiveness of the Cloning attack by double spending in two testnets, one running $\lit{parity}$ and the other running $\lit{geth}$.
%
The application of the Cloning attack to Aura is slower as it consists of the attacker sealing more blocks in one branch while its application to Clique is faster but more subtle as it consists of the attacker  disordering the victim sealers to minimize the weight of a branch.
Overall, we found that Aura required less topological knowledge than Clique for a malicious sealer to achieve double spending with 100\% success rate. 
The attack against Clique is about twice faster than Aura's but its success rate ranges from 60\% to 100\%. 
%
In order to remedy the identified vulnerabilities, we propose to modify these two PoA protocols to preserve their safety.
Even though our counter-measures introduce liveness limitations in these algorithms they make them more suitable for critical applications.

Section~\ref{sec:rw} presents the related work. Section~\ref{sec:model} describes the model. Section~\ref{sec:consensus} formalizes both Aura and Clique protocols as implemented in respectively $\lit{parity}$ and $\lit{geth}$. Section~\ref{sec:attacks} describes the Cloning Attack against both consensus algorithms. 
Sections~\ref{sec:aura-attack} and~\ref{sec:clique-attack} explain how to exploit it to double spend in Aura and Clique, respectively.
Section~\ref{sec:poacore} indicates how we inferred the necessary topological information of the currently running POA Core network.
We then present in Section~\ref{sec:experiments} our evaluation of the Cloning Attack on both protocols, while Section~\ref{sec:analysis} discusses our results and potential countermeasures.  Section~\ref{sec:conclusion} concludes the paper.

\section{Background}\label{sec:rw}

Most of the known \emph{double spending} attacks against blockchains exploit their inherent permissionless 
mechanism by including in the blockchain a transaction that transfers coins and then discards this transaction, hence allowing to re-spend the previously spent coins in a subsequent transaction.
Below we list some of these attacks to explain the recent raise of alternative protocols based on PoA. 

Perhaps the most conventional way to double spend in permissionless blockchains is for an attacker to exploit more than half of the mining power of the system to create a heavier or longer branch that can overwrite transactions that were expected to be sufficiently confirmed or \emph{committed}~\cite{Ros12}.
In some blockchains, a quarter of the mining power appears enough in theory to attract participants into a coalition whose cumulative mining power reaches strictly more than half of the total mining power~\cite{EG14}.
{\sc SmartPool}~\cite{LVTS17} copes with the centralisation of mining power into these blockchains and the risk of mining pools to join a coalition of strictly more than half of the total mining power. 

To attack permissionless blockchains without a significant mining power, researchers attacked
the network. The Eclipse attack against Bitcoin~\cite{HKZG15} consists of isolating at the IP layer a  victim miner from the rest of the network to exploit its resources. The Blockchain Anomaly~\cite{NG16}
exploits message reordering in Ethereum to abort transactions that seemed sufficiently confirmed. The Balance Attack~\cite{NG17} partitions the network into groups of similar mining power to influence the selection of the canonical chain. Recently, actual man-in-the-middle attacks were run to demonstrate the feasibility of stealing assets in Ethereum without a significant mining power~\cite{EGJ18}.

To cope with these attacks, some modern blockchains build upon Byzantine agreement~\cite{PSL80}.
%
sometimes probabilistically~\cite{MSC16,GHM17}, sometimes deterministically~\cite{Kwo15,CNG18}. 
%
Building upon the long Byzantine agreement literature, we know that when the network is \emph{synchronous} and messages are delivered in a known 
bounded time, then $t < \frac{n}{3}$ is sufficient to reach consensus. 
If one also assumes authentication, then even $t < \frac{n}{2}$ becomes sufficient~\cite{LSP82}.
%
What is key for critical applications is that these Byzantine fault tolerant blockchains guarantee that no participants double spend even when messages get unexpectedly delayed. 
Unfortunately, it is impossible to reach consensus when message delays are unbounded~\cite{FLP85}.

Proof-of-Authority (PoA) was recently proposed as a Byzantine fault tolerant consensus mechanism that integrates with the Ethereum protocol~\cite{BBK18}.
The Ethereum $\lit{geth}$ software offers two different PoA consensus protocol, called Clique and Istanbul BFT~\cite{lin_istanbul_nodate} whereas the Ethereum $\lit{parity}$ offers the PoA consensus protocol, called Aura.
The concept is similar to traditional Byzantine fault tolerant consensus in that only $n$ sealers are permissioned to create new blocks but requires authentication and strictly less than $\frac{n}{2}$ Byzantine participants, similarly to the seminal work on Byzantine consensus~\cite{LSP82}.

Although it does not necessarily improve the protocol performance, 
the supposed lack of double spending of PoA raised interest from the industry~\cite{jpmorganchase_quorum,WLT19,BAF19}.
However, some work recently questions the consistency of PoA~\cite{AAB18,shi2018analysis,saltini2019correctness}.
In particular, it was found 
that unsynchronized clocks could affect Aura's consistency whereas Clique was only eventually consistent~\cite{AAB18}, however, no attacks against Aura or Clique have been proposed.
Another work~\cite{shi2018analysis} 
mentioned that an attacker could maintain two chains of equal lengths.
To be possible, this requires the attacker to falsify block timestamps to violate the policy that new blocks are appended to the branch whose latest block has the earliest timestamp among all branches. 
It turns out, however, that such a violation can be easily detected by other sealers verifying timestamp or header of the blocks.
This is probably why the developers did not change the code to remedy this, as the authors mentioned in their paper. 
%
%

Our cloning idea shares similarities with the idea to attack a distributed system by falsifying identities that was already discussed in the past. 
The Sybil Attack~\cite{douceur2002sybil} presents the attack against peer-to-peer systems by forging multiple identities. 
By contrast, the Cloning Attack consists of replicating the machines using the same identity rather than forging identities. More specifically, the Cloning Attack leverages the fact that Ethereum accepts two different machines located at different ends of the network to use the same private key.

\section{Model}\label{sec:model}
We consider a distributed system of $n$ permissioned sealers whose identifiers are $p_1, .., p_n \in \ms{Ids}$. As the blockchain is open, it accepts the requests issued by \emph{nodes} or processes that are not necessarily sealers, hence the overall number of participants can be larger than $n$, but only $n$ participants can propose blocks and \emph{seal} (or sign) them. We assume authentication through a public-key cryptosystem that allows participants to easily identify that a block is correctly signed by a sealer so that incorrectly signed blocks are simply ignored. We assume that keys cannot be forged or stolen by Byzantine participants and that appropriate private keys are correctly distributed to the sealers initially.
As in the Dolev-Yao model~\cite{DY83}, we assume the attacker, who has the control over the Byzantine participants, can intercept messages.

It is claimed that the Aura algorithm is designed to ``tolerate up to 50\% of malicious nodes''~\cite{BBK18}, however, in general a participant cannot decide if half of the nodes pretend that a block is decided while the other half of the nodes pretend that the same block is not decided~\cite{LSP82}. This is the reason why we assume in this paper that no more than $t<\frac{n}{2}$ participants can be \emph{Byzantine} and can act in an arbitrary way, hence a majority of participants always remain \emph{honest}.

As it is well known that consensus cannot be reached in an environment where communication is asynchronous in the presence of faults,
it appears natural to assume additional synchrony. 
It is unclear whether PoA protocols can be safe under \emph{partial synchrony}, where message gets delivered in a bounded amount of time that is not known from the algorithm~\cite{DLS88} or how long communication can take for these protocols to work. 
As an example, a preliminary version of Aura was mentioned to require synchrony in a web document~\cite{noauthor_parity_nodate}, however, this information appears outdated as the implementation is closer to another documentation~\cite{BBK18} that does not mention this assumption, as we explain in Section~\ref{sec:aura}.

The questions we investigate is whether PoA protocols work under partial synchrony, and if not, whether the risk of unexpected message delays is benign (liveness or termination of the consensus is not guaranteed but safety remains guaranteed in that no double spending occur) or can have dramatic consequences (disagreement can occur), hence letting an attacker double spend.
As we will explain, our conclusion is that PoA protocols do not work in that even safety is not guaranteed but some countermeasures can remedy this problem.

\section{PoA Consensus Algorithms}
\label{sec:consensus}

In this section we describe the two main variants of PoA algorithms, called Aura and Clique, implemented in the predominant Ethereum software, called $\lit{parity}$ and $\lit{geth}$, respectively.
We first formalize two distinct versions of the Aura algorithm that are both publicly available online. We then formalize the Clique algorithm.

\subsection{The Aura consensus algorithms}\label{sec:aura}

There exist two distinct versions of the Aura algorithm as documented online, one that corresponds to the current $\lit{parity}$ implementation of the Ethereum protocol and another~\cite{BBK18} that uses rounds to decide whether a consensus decision is reached.

\subsubsection{The $\lit{parity}$ Aura algorithm}
Algorithm~\ref{algo:aura} depicts the way Aura guarantees that participating nodes reach consensus on the uniqueness of the block at a given index of the blockchain as implemented in 
$\lit{Parity-Ethereum-v2.0.8}$ (v2.0.8 was the latest version at the time we performed our experiments).
Every participating node maintains a state comprising a set of $\ms{sealers}$, its current view of the blockchain $c_i$ as a directed acyclic graph $\langle B_i, P_i \rangle$, a block $b$ with fields $\ms{parent}$ that links to the parent block, a $\ms{sealer}$ and a $\ms{step}$ indicating the time at which the block is added to the blockchain, as explained below. Initially, they are $\bot$ meaning ``undefined''.

\begin{algorithm}[ht!]
	\caption{The $\lit{parity}$ Aura algorithm at process $p_i$}
	\label{algo:aura}
	\begin{algorithmic}[1]
		{\footnotesize                  
			
			\Part{State}{
				\State $\ms{sealers} \subseteq \ms{Ids}$, the set of sealers
				\State $\ms{c_i} = \langle B_i, P_i \rangle$, 
				the local blockchain at node $\ms{p_i}$ 
				is a directed 
				\State \T acyclic graph of blocks $\ms{B_i}$ 
				and pointers $\ms{P_i}$
				\State $\ms{b}$, a block record with fields:
				\State \T$\ms{parent}$, the block preceding $\ms{b}$ in the chain, initially $\bot$
				\State \T$\ms{sealer}$, the sealer that signed block $b$, initially $\bot$
				\State \T$\ms{step}$, the blockchain step when the block gets added, initially $\bot$
				\State $\lit{step-duration}$, the duration of each step as configured
			}\EndPart
			
			\State
			
			\Part{$\lit{propose}()_i$}{ \label{part:propose} \Comment{sealers keep proposing}
				\While{$\lit{true}$}		\Comment{infinite loop}
				\State $\ms{step} \gets \lit{clock-time()} / \lit{step-duration}$ \label{line:clock-time} \Comment{discretize time}
				\If{$i \in \ms{sealers} \wedge \ms{step} \lit{~mod~} |\ms{sealers}| = i)$} \label{line:turn} \Comment{my turn}
				\State $\ms{b.parent} \gets \lit{last-block}(c_i)$ \Comment{link a block}
				\State $\ms{b.sealer} \gets \ms{p_i}$ \label{line:seal} \Comment{seal the block}
				\State $c_i \gets \langle B_i\cup\{b\}, P_i\cup\{b.\ms{parent}\} \rangle$ \Comment{update local view}
				\State $\lit{broadcast}(c_i)$ \Comment{send blocks}
				\EndIf
				\State $\lit{sleep}(\lit{step-duration})$ \Comment{wait before looping}
				\EndWhile
			}\EndPart
			
			\State 
			
			\Part{$\lit{score}(\langle B_j, P_j \rangle)$}{ \Comment{compute the score of a branch}  
				\label{part:score}
				\Return $\lit{UINT128\_MAX} \times \lit{height}(\langle B_j, P_j \rangle) - \lit{step-num}(\langle B_j, P_j \rangle)$
				\EndReturn
			}\EndPart
			
			\State
			
			\Part{$\lit{deliver}(\langle B_j, P_j \rangle)_i$}{   
				\label{part:deliver}
				\If{$\lit{score}(\langle B_j, P_j \rangle ) > \lit{score}(\langle B_i, P_i \rangle )$}
				\State $\langle B_i, P_i \rangle \gets \langle B_j, P_j \rangle$	\Comment{select the right branch in case of forks} 
				\EndIf
			}\EndPart
			
			\State
			
			
			\Part{$\lit{is-decided}(b)_i$}{ \label{line:decided}
				\State $\ms{V} \gets \{ \ms{b_k.sealer} \;|\; \ms{b_k} \in B_i ; k \geq i\}$ \Comment{sealers in blocks since $b_i$}
				\Return $(|\ms{V}| \times 2 > |\ms{sealers}|)$ \Comment{more than majority of sealers signed} \label{line:aura-is-decided}
				\EndReturn
			}\EndPart
			
		}
	\end{algorithmic}
\end{algorithm}

The function $\lit{propose()}$ is invoked in order to propose a block for a particular index of 
the blockchain. The consensus is reached once the block is decided, which can happen much later as we will explain in the function $\lit{is-decided()}$ (line~\ref{line:decided}) below. The algorithm discretises time into steps that corresponds to consecutive periods of 
$\lit{step-duration}$ time, as specified in a configuration file. 
Each sealer executes an infinite loop that periodically checks whether the $\lit{clock-time()}$ indicates that this is its turn to propose a block (line~\ref{line:clock-time}).
When it is its turn (line~\ref{line:turn}), a sealer sets  the parent of the block to the last block of its view and signs it (line~\ref{line:seal}).

Each $\lit{broadcast()}$ invoked by the $\lit{propose()}$ function sends blocks that get delivered
to all other participating nodes that are honest (in reality only the last block is broadcast unless some sealer is unaware of more blocks). The $\lit{deliver()}$ function (line~\ref{part:deliver}) is thus invoked at each honest participating node, regardless of whether it is a sealer, upon reception of the broadcast message.
Once a blockchain view is delivered to $p_i$, the node compares the score of the blockchain view it maintains to the blockchain view it receives, using the $\lit{score}$ (line~\ref{part:score}). 
The highest blockchain has the greatest score, however, if two blockchains share the same height, then the one that is denser in terms of its number of non-empty slots obtains the highest score.  This is indicated by the two functions $\lit{height}$ and $\lit{step-num}$ that represent the height of the blockchain and the number of slots for which there exists a block in the blockchain.

\subsubsection{Round-based variant of the Aura algorithm}\label{sec:aura-alt}

The Aura algorithm implemented in $\lit{parity}$ is not the only algorithm called, Aura.
Another variant is presented in the PoA Network white paper available online~\cite{BBK18}.
Algorithm~\ref{algo:aura-alt} presents the different decision technique of this variant, the rest of the pseudocode being identical to Algorithm~\ref{algo:aura}.

In order to know whether a block $b$ is \emph{decided} at the end of a successful consensus (Algo~\ref{algo:aura-alt}, line~\ref{line:alt-decided}), a participant simply has to check whether there exist two consecutive rounds $\ms{round1}$ and $\ms{round2}$ following block $b$, in each of which the blocks are sealed by a majority of the sealers.

\begin{algorithm}[ht!]
	\caption{The round-based variant of Aura at process $p_i$}
	\label{algo:aura-alt}
	\begin{algorithmic}[1]
		{\footnotesize

			\Part{$\lit{is-decided}(b)_i$}{ \label{line:alt-decided}
				\State $\ell \gets |\ms{sealers}|$ \Comment{number of validators}
				\State $\ms{round1} \gets (\lit{b.step}, \lit{b.step}+\ell]$ \Comment{steps in next round}
				\State $\ms{round2} \gets (\lit{b.step}+\ell, \lit{b.step}+2*\ell]$ \Comment{steps in the 2nd next round}
				\State $\ms{maj1} \gets |\{b': \ms{b'.step} \in \ms{round1}\}| > \ell / 2$ \Comment{majority in round 1}
				\State $\ms{maj2} \gets |\{b'': \ms{b''.step} \in \ms{round2}\}| > \ell / 2$ \Comment{majority in round 1}				
				\Return $(\ms{maj1} \wedge \ms{maj2})$ \Comment{decided if majority in both rounds}
				\EndReturn
			}\EndPart
		}
	\end{algorithmic}
\end{algorithm}

Note that while presented in some documentation, this alternative Aura specification is not the one used by the mainstream implementation of the protocol.
The current definition of the Aura algorithm disregards the rounds and simply requires enough blocks to be sealed~\cite{noauthor_parity_nodate}.
Although the version of Aura we experiment in this paper is the mainstream one (Algorithm~\ref{algo:aura}), the attack we present in Section~\ref{sec:aura-attack} also applies to this more restrictive definition presented in Algorithm~\ref{algo:aura-alt}.

\subsection{The Clique consensus algorithm}\label{sec:clique}

\sloppy{Algorithm~\ref{algo:clique} depicts the pseudocode of the Clique consensus algorithm.
It is the one used currently in $\lit{geth-1.8.20-stable}$.}

Every participating node shares the same initial block, the genesis block, which also contains the $\lit{block-period}$, the period between consecutive block creations.
Similarly to the Aura protocol, each node maintains its own view of the growing blockchain $c_i$ as a directed acyclic graph $\langle B_i, P_i\rangle$.
A block $b$ contains a $\ms{number}$ as an index of the block in the blockchain, a $\ms{weight}$ as a weight of the block, a $\ms{parent}$ field that links to its parent block and a $\ms{sealer}$.

\begin{algorithm}[ht!]
	\caption{The $\lit{geth}$ Clique algorithm at process $p_i$}
	\label{algo:clique}
	\begin{algorithmic}[1]
		{\footnotesize                  
			
			\Part{State}{
				\State $\ms{sealers} \subseteq \ms{Ids}$, the set of sealers
				\State $\ms{c_i} = \langle B_i, P_i \rangle$, 
				the local blockchain at node $\ms{p_i}$ 
				is a directed 
				\State \T acyclic graph of blocks $\ms{B_i}$ 
				and pointers $\ms{P_i}$
				\State $\ms{b}$, a block record with fields:
				\State \T$\ms{parent}$, the block preceding $\ms{b}$ in the chain, initially $\bot$
				\State \T$\ms{sealer}$, the sealer that signed block $b$, initially $\bot$
				\State \T$\ms{number}$, the index of the block in the chain, initially $\bot$
				\State \T$\ms{weight}$, the weight of a block, initially $\bot$
				\State $\lit{block-period}$, minimum duration in second between timestamps of 
				\State \T two consecutive blocks, initially $5$ seconds
				\State $\lit{majority}$, the number of  $\lfloor\frac{|\ms{sealers}|}{2}\rfloor+1$
				\State $\lit{sealer-limit}$, max. number of consecutive blocks among which a sealer   
				\State \T  can sign at most one block, initially set to the $\lit{majority}$ \label{line:clique-sealer-limit}
			}\EndPart
			
			\State
			
			\Part{$\lit{sign-recently}(c_i, n)_i$}{
				\label{part:sign-recently}
				\State $\ms{\lambda} \gets \lit{sealer-limit}$
				\State $\ms{ret} = \lit{false}$
				\For{$\ms{m} = n - \lambda, ..., n$} \Comment{loop through last $\lambda$ blocks}
					\T\If{$\ms{b_m}.\ms{number} \lit{~mod~} |\ms{sealers}| = i$}
					 $\ms{ret} = \lit{true}$
					\EndIf
				\EndFor
				\Return $\ms{ret}$
				\EndReturn
			}\EndPart			
			
			\State
			
			\Part{$\lit{propose}()_i$}{
				\label{part:propose}
				\While{$\lit{true}$}
				\State $\ms{n} \gets \lit{last-block}(c_i).\ms{number}$ \Comment{last block index}
				\WUntil{$\neg \lit{sign-recently}(c_i, n)$} \label{line:clique-check-sign-recently} \Comment{wait until I can seal a block}
				\EndWUntil
				\State $\ms{T} \gets \lit{get-last-timestamp}(c_i)$ 
				\WUntil{$\lit{clock} \geq T+\lit{block-period}$} \label{line:clique-check-block-period}\Comment{wait $\geq$ block-period}
				\EndWUntil
				\If{$(\ms{n+1}) \lit{~mod~} |\ms{sealers}| = i$} \label{line:clique-check-in-turn} \Comment{in-order sealing}
					\State $\ms{b.weight} = 2$ \Comment{block weight 2}
				\Else \Comment{out-of-order sealing}
					\State $\lit{sleep}(\lit{rand}([0, 500 \times \lit{majority}])\,\ms{ms})$ \Comment{random delay in millisecs} \label{line:delay}
					\State $\ms{b.weight} = 1$	\Comment{block weight 1}
				\EndIf
				\State $\ms{b.number} = \ms{n} + 1$ \Comment{increment block index}
				\State $\ms{b.parent} \gets \lit{last-block}(c_i)$ \Comment{link a block}
				\State $\ms{b.sealer} \gets \lit{sign()}$ \label{line:seal} \Comment{seal the block}
				\State $c_i \gets \langle B_i\cup\{b\}, P_i\cup\{b.\ms{parent}\} \rangle$ \Comment{update local view}
				\State $\lit{broadcast}(c_i)$ \Comment{send blocks} 

			    \EndWhile
			}\EndPart
			
			\State
			
			\Part{$\lit{total-weight}(\langle B_j, P_j \rangle )_i$}{ \label{part:clique-weight} \Comment{total weight}
				\Return $\sum \ms{b.weight} | \ms{b} \in \ms{B_j}$
				\EndReturn
			}\EndPart
			
			\State 
			
			\Part{$\lit{deliver}(\langle B_j, P_j \rangle)_i$}{ 
				\label{part:clique-deliver}
				\If{$\lit{total-weight}(\langle B_j, P_j \rangle ) > \lit{total-weight}(\langle B_i, P_i \rangle )$} \Comment{heaviest}\label{line:clique-heavier}
				\State $\langle B_i, P_i \rangle \gets \langle B_j, P_j \rangle$	
				\EndIf
			}\EndPart
			
			\State
			
			\Part{$\lit{is-decided}(b)_i$}{ \label{line:clique-decided}
				\State $\ms{V} \gets \{ \ms{b_k.sealer} : \ms{b_k} \in B ; k \geq i\}$ \Comment{sealers in blocks since $b_i$}
				\Return $|\ms{V}|  > \lit{majority}$ \Comment{more than majority of sealers signed}\label{line:clique-decides}
				\EndReturn
			}\EndPart
			
		}		
	\end{algorithmic}
\end{algorithm}

The $\lit{propose()}$ function runs an infinite loop in order to propose blocks to the blockchain when  certain conditions are satisfied.
The first condition (line~\ref{line:clique-check-sign-recently}) requires the process to wait for blocks from other sealers until none of the last $\lit{sealer-limit}$ blocks contains its signature. 
In the current implementation the $\lit{sealer-limit}$ must be $\lfloor|\ms{sealers}| / 2\rfloor + 1$, which is the smallest $\lit{majority}$.
As a result of this first condition, sealers need to take turn to sign blocks.
The second condition (line~\ref{line:clique-check-block-period}) is 
to wait for $\lit{block-period}$.\footnote{The default $\lit{block-period}$ is 15 seconds as developers suggest the same duration to remain analogous to the proof-of-work blockchain Ethereum.}
%
When both conditions are met, the process checks if it is its turn to sign the block (line~\ref{line:clique-check-in-turn}).
The process may sign a block right away with $\ms{weight}$ equal to 2; otherwise, it may sign a block with $\ms{weight}$ equal to 1 after a random delay between 0 and $500 \times \lfloor|\ms{sealers}| / 2\rfloor + 1$ milliseconds (line~\ref{line:delay}).
The consensus is reached once the block is decided later as we will describe in the function $\lit{is-decided()}$ (line~\ref{line:clique-decided}).
The last step in the loop, $\lit{broadcast()}$, sends messages to other participants.

Upon reception of the broadcast message, the $\lit{deliver()}$ function (line~\ref{part:clique-deliver}) is invoked at each participating node regardless of whether it is a sealer.
The $\lit{total-weight}$ function (line~\ref{part:clique-weight}) used by the process compares the weight between two blockchain views, a current blockchain that it maintains locally and the one freshly received.
The process updates its local view if the received blockchain is heavier; otherwise it keeps the same local blockchain view.

To consider whether a block $b$ is \emph{decided} (line~\ref{line:clique-decided}), a process has to check the set of sealers who sign blocks after $b$. Only when a majority of sealers have appended subsequent blocks to the chain, can a block be considered decided.


\section{The Cloning Attack}
\label{sec:attacks}

In this section, we present the cloning attack to double spend in PoA blockchains. 
In particular, we present the commonalities between the attacks against Aura and Clique, namely the cloning process that allows an attacker to play different roles in the blockchain, the majority that allows two groups of sealers to make progress without the other, and the way transactions should conflict to double spend.
The difference in how these attacks are applied to Aura and Clique are deferred to Sections~\ref{sec:aura-attack} and~\ref{sec:clique-attack}, respectively.

By assumption, only a minority of the sealers can be malicious, this is the reason why PoA algorithms require a majority of sealed blocks to consider whether a block and its transactions appear to be committed. Intuitively, this should prevent the malicious sealers to form a coalition that can double spend. In reality, as we explain below, $(2 - (n\;\text{mod}\;2))$ attacker(s) cloning their own instance into two clones are sufficient to double spend.

\subsection{Cloning instances by duplicating keys}

The first step necessary in the Cloning Attack is for some attacker to duplicate its Ethereum instance into two clones.
\emph{Cloning} consists for a single user of running two instances of the Ethereum protocol with the same \emph{address} or public-private key pair.  Note that these two instances could run either on the same machine, using the same IP address, or on distinct machines with distinct IP addresses. We call these two instances \emph{clones} because one has the same information as the other before messages start being delayed. In addition, during the whole duration of the attack, both clones use the same public-private key pair. 
Interestingly, we noted that Ethereum allows these two cloned instances to both create blocks, however, as they use the same private key to seal blocks, they are considered to act as a unique sealer.

At some point, the attacker exploits message delays (either accidental or as a result of a network attack) between two groups of a minority of $\lceil n/2 \rceil - 1$ sealers, hence creating a transient partition. At this moment, the two clones may not share  exactly the same database content as they may not be aware of the exact same blocks that are present in the blockchain. To maintain the cloning at the start of the partition, the attacker copies the content of the blockchain database of one of the clones to the database of the other clone and connects each of these clones to a different partition. During the time of this partition, the Ethereum protocol readjusts the peering so that sealers within the same group keep communicating.

Note that they are various ways of obtaining a partition in the Ethereum network either by  misconfiguring routes, leveraging natural disasters or maliciously attacking the network. 
One example of a network attack that allows to partition the Internet is the BGP hijacking attack. It works by having an attacker advertising to one group wrong routes that reach the other group in order to intercept all traffic between the two groups. Once the traffic is rerouted, the attacker can simply delay the propagation of messages.
We refer the interested reader to existing ways of implementing man-in-the-middle attacks in Ethereum~\cite{EGJ18}. 

\subsection{Majority groups to guarantee progress}
Clones are exploited in the attack to give the illusion to honest sealers that each group contains a majority of sealers.
In order to progress towards a double spending situation, each group must commit transactions and thus decide blocks, this is why we need  $(2 - (n\;\text{mod}\;2))$ attackers that clone instances:
\begin{itemize}
\item {\bf Case $n$ is odd.} The honest sealers can be split into two groups of $(n-1)/2$ sealers, each representing a minority. In order to guarantee progress of the protocol on both sides of the partition, a single attacker can simply add one clone in each minority, hence reaching a majority of $\lfloor n/2 \rfloor +1$ sealers on each side.  This is the reason why $(2 - (n\;\text{mod}\;2))=1$ attacker is sufficient when $n$ is odd.
\item {\bf Case $n$ is even.} A single attacker could split the $n-1$ honest sealers into two groups of different sizes, one that contains $n/2$ sealers and another that contains $n/2-1$ sealers. 
It would however be insufficient to include a clone in the second group to guarantee its progress.
This is why $(2 - (n\;\text{mod}\;2))=2$ attackers are needed.
\end{itemize}

To conclude, the $(2 - (n\;\text{mod}\;2))$  attacker(s) thus partition(s) a network of $n$ sealers into roughly two halves to which they add clones so that each group contains a majority of at least $\lfloor n/2 \rfloor +1$ sealers. This guarantees the progress of the protocol on each group so as to obtain the commit of a transaction $\ms{TX_1}$ on one group and the canonical chain containing $\ms{TX_2}$  in the other group.
%
For example, there must be at least 5 sealers in each subgroup for a network of $n=9$ sealers.
Such a condition is required to ensure termination of the consensus algorithm, so that blocks will be decided, or appear to be final, from the viewpoint of both subgroups.
%

Note that we consider here the necessary time for a partition. In a realistic scenario, the attacker may want the effect of its transaction to take occur before stopping the partition. For example, an attacker buying a good in transaction $\ms{TX_1}$ may want to receive the good before the transaction gets discarded from the blockchain.

\subsection{Conflicting transactions}
The most common way of executing a double spending is to make sure a transaction $\ms{TX}_1$ ends up being included in one branch of a fork, then convincing the recipient that $\ms{TX}_1$ is committed, before resolving the fork by discarding the branch of this transaction $\ms{TX}_1$. Later on, the sender of the transaction $\ms{TX}_1$ can simply reuse the coins he initially spent in $\ms{TX}_1$ in another transaction $\ms{TX}_2$.
Interestingly in Ethereum, if the conflicting transaction $\ms{TX}_2$ is not issued early enough, then 
 $\ms{TX}_1$ could be re-included in a mempool and committed later on.
 
The goal is for the clones to leverage the message delays between network partitions to rapidly issue two conflicting transactions.
As soon as the blockchain network is divided into two subgroups, the attacker issues a minimum of two conflicting transactions, at least one transaction to each subgroup.
A typical example to illustrate the double spending attack is two conflicting transactions:   
\begin{enumerate}[leftmargin=.4in]
\item[$\ms{TX}_1$] where Alice gives all her coins to Bob in the first transaction sent to one group and 
\item[$\ms{TX}_2$] where Alice gives all her coins to Carol on the other transaction sent to the other group. 
\end{enumerate}
It is clear that committing both transactions would violate the integrity of Alice's account and would result in a double spending.
Once the first transaction appears committed, delivering the delayed messages or ending the partition will have the effect of discarding one of the two transactions. 

In the next two sections, we explain how the majority of sealers in Aura and the order of the sealings in Clique allow to select the transaction to be discarded by the system.

%
%

\section{The Cloning Attack Against Aura}\label{sec:aura-attack}

%
We now present a simple way to apply the Cloning attack to double spend in Aura.
To discard the branch, say the \emph{victim branch}, that contains $\ms{TX_1}$ and double spend, the attacker must influence Aura to select the branch containing $\ms{TX_2}$, say the \emph{attacker branch}, as the canonical chain.

As explained earlier in Algorithm~\ref{algo:aura}, the current implementation of Aura
simply chooses the longest chain as the canonical chain whenever a fork is detected.
So, to influence the selection of the attacker branch as the canonical chain, the attacker simply has to contribute to the attacker branch by sealing more blocks in the group maintaining this branch than the other group.
%
%
%


\begin{figure}[ht!]
	\centering
	\includegraphics[width=.5\textwidth]{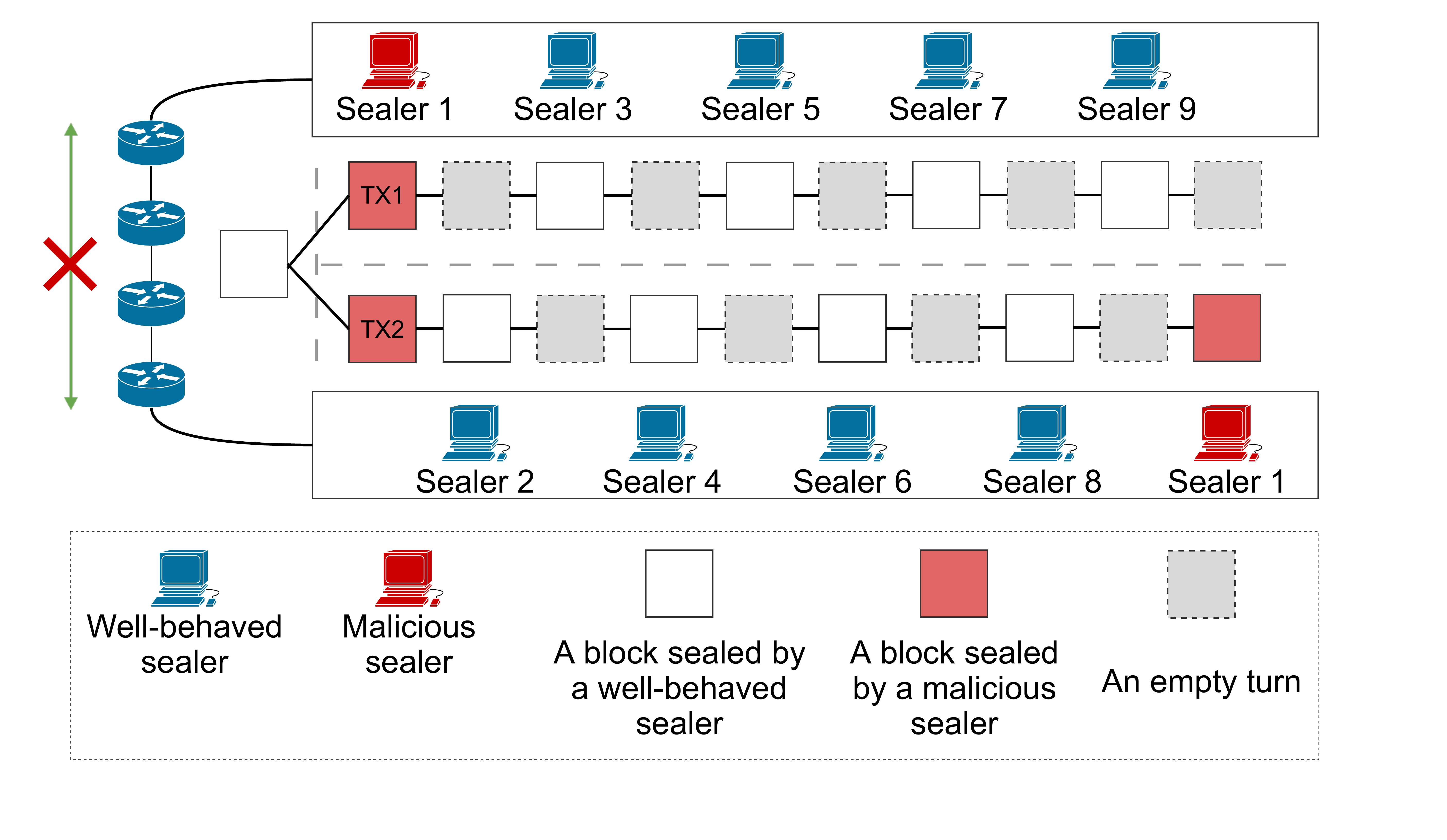}
	\caption{Applying the cloning attack to double spend in Aura requires 
	the attacker, ``Sealer 1'', to delay messages during $(n+1)\times s$ seconds for transaction $\ms{TX_1}$ to be committed on the upper branch and for the attacker to seal more blocks on the lower branch than on the upper branch}
	\label{fig:attack-poa}
\end{figure}


%


To seal more blocks in one branch than another, the attacker maintains the partition during $(n+1) \times s$ seconds, where $n$ is the number of sealers and $s$ is the step duration in seconds that separates consecutive blocks.
The reason is twofold. 

\begin{itemize}
\item First, as mentioned earlier in Algorithm~\ref{algo:aura}, Aura requires $ns$ delay after a block is created to ensure that it is decided. Deciding a block on the victim side is necessary to make sure that $\ms{TX_1}$ gets committed. Given that both the size of the group on each side is $\lfloor n/2 \rfloor + 1$ and that each sealer seals one after another, the attacker clone must also seal at least one block. 
%
%
%
%
%
%
%
\item Second, the attacker must ensure that the attacker branch is longer than the victim branch so that the attacker branch gets selected by Aura as the canonical branch. This can only be done if the attacker seals two blocks on the attacker branch, i.e., one extra block compared to the number of blocks it sealed on the victim side. As a result, the attacker needs to maintain the network partition for $(n + 1)\times s$ seconds to get at least two turns in which it can seal a block. 
\end{itemize}

\paragraph{Example with 9 sealers} For the sake of simplicity, \figurename~\ref{fig:attack-poa} depicts the cloning attack against Aura with a network partition where there are $n=9$ sealers and where $2 - n\;\text{mod}\;2=1$ sealer is malicious, namely ``Sealer 1''. 
This attacker is thus present in both groups through its two cloned instances and gives the illusion that each group contains a majority of $\lfloor n/2\rfloor  + 1 = 5$ sealers while one of the sealers in each group is actually a clone.
%
As we can see, this attack translates into having Sealer~1 creating the last block (depicted with the red right-most block in the figure) only on the lower partition before merging the two partitions. 
By doing so, Sealer~1 makes sure that this branch will be the canonical branch whereas the upper branch will disappear. The attacker is thus guaranteed to double spend successfully.

\section{The Cloning Attack Against Clique}\label{sec:clique-attack}

In this section, we apply the cloning attack against Clique.
%
In Clique, the Cloning Attack does not require to take as long as in Aura.
Unlike in Aura, a sealer of Clique can seal a block even when it is not its turn. Depending on their turn, some sealers may have to wait while others do not. 
These differences impact the way the attacker can influence the selection of one branch of a fork as the canonical chain and allow an attacker to double spend faster than in Aura.

\subsection{In-order and out-of-order sealers}
The cloning attack against Clique differs from the one against Aura 
in the moment at which it starts delaying messages.
Because the order of sealing is important in Clique, the attacker should ideally decide to start delaying the messages based on the sealer's turn to seal a block.

When a sealer seals a block while it is his turn, we call this sealer an \emph{in-order sealer} and the block an \emph{in-order block} (cf. Alg.~\ref{algo:clique}, Line~\ref{line:clique-check-in-turn}). There is at most one in-order sealer to seal the current block in each partition 
of Clique.
When a sealer seals a block while it is not his turn, we call this sealer an \emph{out-of-order sealer} and this block an \emph{out-of-order block} (cf. Alg.~\ref{algo:clique}, Line~\ref{line:delay}).
As a sealer must wait for $\ms{sealer-limit}$ blocks between two blocks it seals, there are at most $(n-\ms{sealer-limit})$ potential out-of-order sealers to seal a block.
The in-order block contributes a weight of 2 to the weight of its branch whereas the out-of-order blocks contribute to 1 to the weight of its branch, hence sealing in-order or out-of-order impacts the decision regarding the branch selection process.

In addition, an in-order sealer can append a block to the chain without waiting for any delay as
shown in Line~\ref{line:clique-check-in-turn} of Alg.~\ref{algo:clique}. By contrast, an out-of-order sealer has to wait for a random period as indicated at Line~\ref{line:delay} of Alg.~\ref{algo:clique}. This mechanism gives the in-order sealer some time to be the first to seal a block in his turn, but allows out-of-order sealers to seal a block if the in-order sealer is lagging.

As the canonical chain is chosen among the branches of a fork by comparing the sum of their block weights, the attacker must have a maximum number of in-order sealers at the time of the partition to maximize the overall weight. 
Hence, to influence the selection of the branch as the canonical chain, the attacker must choose the 
proper turn to start delaying messages. If not done properly, the attacker risks to maximizing 
the weight of the branch where its transaction was included, limiting the chances of a successful double spending.

\begin{figure}[t]
	\centering
	\includegraphics[width=.5\textwidth]{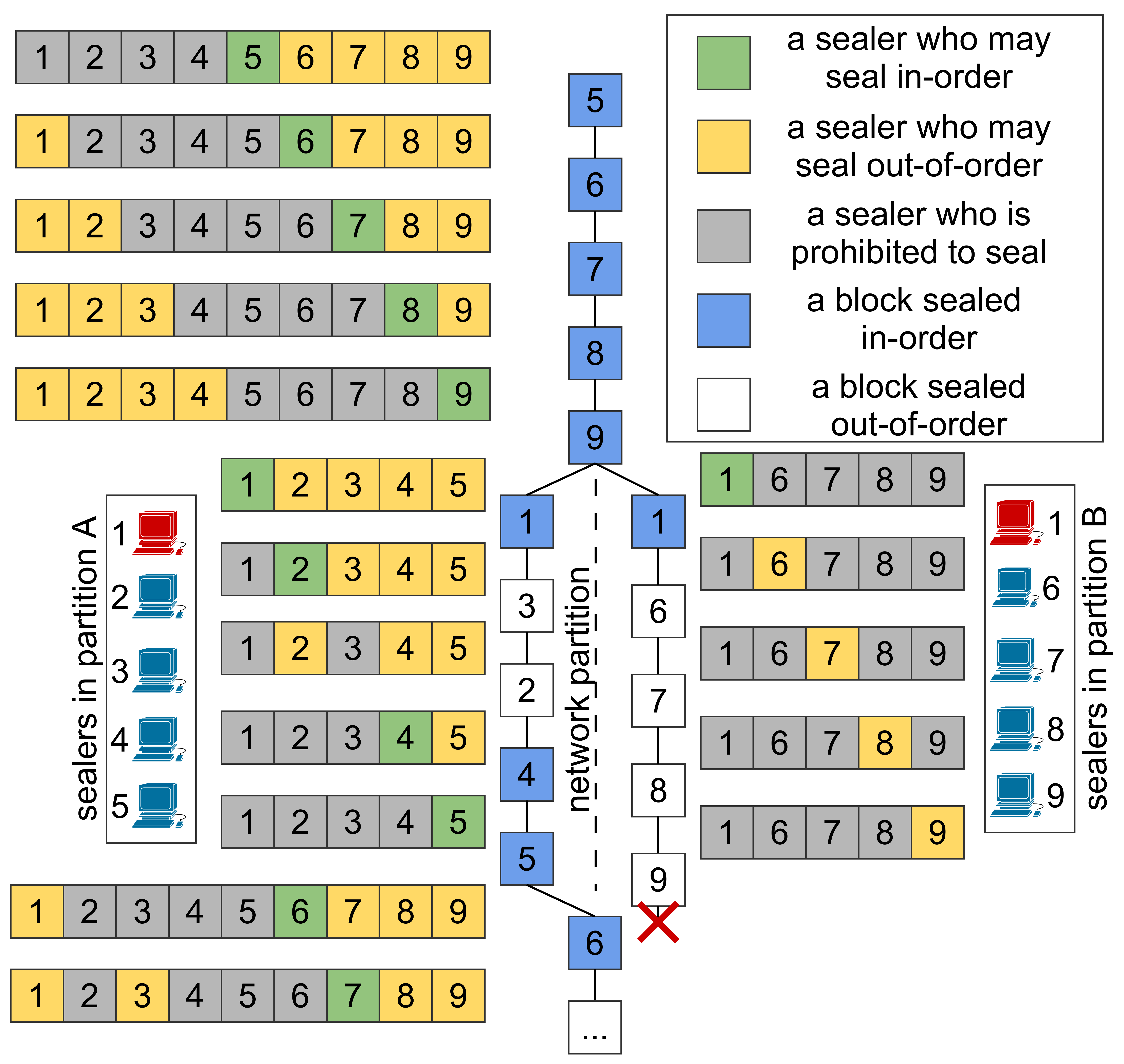}
	\caption{An execution  of the in-order cloning attack  against Clique where $n=9$ sealers mine the blue blocks in-turn before the messages get delayed, after which each group seals five blocks, 3 in-order blocks on the left group and 1 in-order block on the right group}
	\label{fig:clique_attack}
\end{figure}


\subsection{Disordering sealers to select a branch}



\figurename~\ref{fig:clique_attack} depicts the execution of the attack with $n=9$ sealers and one attacker (Sealer 1) as time increases from top to bottom.
Initially, the blockchain starts with block 5, indicating that the first block is sealed by Sealer 5. 
As times goes on, Sealers 6, 7, 8 and 9, seal one after the other the subsequent blocks of the blockchain. As there is no partition yet, the in-order sealers are the first to sign these blocks during their respective turn, hence all blocks are in-order blocks represented in blue in the figure.
Next to each created block is a list of sealers that are either unable to seal (grey), in-order sealers (green) or out-of-order sealers (yellow).

Consider that Sealer 1, the attacker, performs the cloning and delays the network messages. 
Right after Sealer 9 sealed his block, Sealer 1 starts intercepting the messages between the group of sealers 2, 3, 4 and 5 on the left side and the group of sealers 6, 7, 8 and 9 on the right side. 
Note that Sealer 1 is represented on both sides because of the presence of one of its clones on each side. The resulting partition is indicated in 
\figurename~\ref{fig:clique_attack} with a fork of the blockchain into two branches. 
Right after the partition starts, Sealer 1 issues two conflicting transactions $\ms{TX_1}$ and $\ms{TX_2}$
on each side of the partition that will double spend.
The two clones of Sealer 1 
allow him to seal one block in each group. Note that these blocks are labelled 1 and represented in blue because Sealer 1 is the in-order sealer at this point in time.  
After sealing, Sealer 1 is no longer able to seal any block due to the $\ms{sealer-limit}$, hence Sealer 1 is depicted in grey in both groups.

On the right side of the partition, we can see that Sealer 6 seals the following block, even though  it is not the in-order sealer at this moment. This is because the in-order sealer, Sealer 2, cannot communicate with this group as the network is partitioned.
For some reason it might also be the case  on the left side of the partition that Sealer 2 is not fast enough to seal the next block and that another sealer, say Sealer 3, manages to seal it before. Note that this can happen as the delay 
Sealer 3 has to wait before sealing is a random number that can be null (cf. Alg.~\ref{algo:clique}, Line~\ref{line:delay}). 
However, this last seal from Sealer 3 prevents it from sealing the next block in-order as it has to wait for the $\ms{sealer-limit}$, hence the next block is again out-of-order. 
The process continues where sealers on the left side seal in-order whereas sealers on the right side seal out-of-order.

Finally, the attacker does no longer need to delay the messages and can stop the partition
as both transactions $\ms{TX_1}$ and $\ms{TX_2}$ are now successfully committed. In fact, the transactions are both now in the first block of a series of $\lfloor n/2 \rfloor + 1 = 5$ consecutive blocks,  which is sufficient for all Clique users to consider these transactions as committed because their block is decided as indicated at Line~\ref{line:clique-decides} of Algorithm~\ref{algo:clique}.
We can conclude that the weight gained by the branch on the left side during the partition is $3\times2 + 2\times1= 8$ because it contains 3 in-order blocks and 2 out-of-order blocks.
By contrast, the weight gained by the branch on the right side during the partition is $1\times2 + 4\times1= 6$ because it contains 1 in-order block and 4 out-of-order blocks. 
It follows from the difference in weight of the two branches that the heaviest branch on the left side is chosen as the canonical branch whereas the lightest branch on the right side is simply discarded by the protocol (cf. Alg.~\ref{algo:clique}, Line~\ref{line:clique-heavier}).

%
%

\subsection{Attack regardless of the order of sealers}
Note that even if the attacker does not know the topology, there is a way to attack Clique. The attack is slightly different from the previous one as it relies on the possibility for the attacker to become the only sealer able to seal a block on both sides of the partition.
The attacker can simply seal a single block on the victim branch, and keep sealing blocks on the attacker branch. In the worst case scenario for the attacker, all the $\lfloor n/2 \rfloor +1$ upcoming in-order sealers end up on the victim side, which will maximize the weight of the branch on the victim side gained during the partition. 
Recall that the $\ms{sealer-limit}$ is always $\lfloor n/2 \rfloor +1$ in Clique (Alg.~\ref{algo:clique}, Line~\ref{line:clique-sealer-limit}),
Now, if the attacker stops sealing a second block on the victim side, then the maximum weight gained on this side during the partition will be $(\ms{sealer-limit}\times 2)$. 
The attacker simply needs to keep sealing on the other branch until the gained weight on this branch  reaches $(\ms{sealer-limit}\times 2 + 1)$. 
In this case, the attacker successfully double spends regardless of the sealer turn in each group.
 %

\section{Inferring the POA Core Topology}\label{sec:poacore}

In this section, we show how to infer the topological information necessary to run the Cloning attack against the currently running POA Core network. 

Since POA Core is a public blockchain, the network requests its sealers to comply with some eligibility requirements as part of an application procedure.
First, all individuals running a sealer node must be US residents. Second, these individuals must deploy their Ethereum sealer instance within the US. Third, a sealer candidate must be a public notary in the US with a valid license. (Note that this is not a hard requirement for Sokol, but it is mandatory for POA Core.) 
Fourth, sealer individuals must write, in a POA forum, posts about themselves that often include their full name and their notary public licenses. (This can then be used to retrieve their mail addresses.)

To bootstrap the POA Core network, the first sealer of a network called the ``Master of Ceremony'' generates the initial keys and distributes them to a group of independent participants. Together they form the first group of sealers that govern the blockchain network.
This governance may vote for adding new sealers or removing existing ones to limit, for example,  bribery attacks. 
Before a new sealer candidate can be eligible for a sealer role on POA Core, one needs to apply a sealer role on Sokol, a testnet for POA Core, and actively participate in its on-chain governance for some minimum period of time.

\paragraph{Netstat information}
In addition to the public information part of the sealer application, POA Core employs two monitoring tools that also publicizes information about its sealers.
For the public to be able to browse blockchain information conveniently, POA Core has its own block explorer site~\cite{noauthor_poacore_nodate}, which reveals the sealing order of all the active sealers as well as the required time to seal one block.
In order to monitor the system and the network status of each sealer, POA Core incorporates the $\lit{netstat}$ server within its own web-based dashboard site~\cite{noauthor_poacore_netstat_nodate} and displays the information received from all participating sealers.
Some of this information, including the number of peers of each sealer, its OS names and its latency allowed us to uniquely identify the sealer nodes.

\paragraph{Active measurement and fingerprinting}
In order to obtain the versions of the parity protocol run by sealers, their IP addresses, and their port numbers, we instrumented the code of Ethereum $\lit{parity}$ and launched a new node participating in POA Core.
Using the information gathered from the public sources, we inferred the topology of the POA Core network to delay message for the Cloning attack.
We used the OS names obtained from the $\lit{netstat}$ server dashboard in order to guess the datacenters where the nodes are deployed.
Assuming the sealers reported correct information, the suffix of the OS names reported to the $\lit{netstat}$ server indicates the cloud provider where the node runs, for example $\lit{aws}$ indicates Amazon Web Service while $\lit{azure}$ indicates Microsoft Azure.
The information gathered by our node from its peers included non-sealer nodes such as boot nodes, so we eliminated some of these non-sealer nodes by examining their port numbers and software versions.
For instance, it is trivial that a node that runs Parity on $\lit{tcp/21000}$ is not a sealer, as the $\lit{netstat}$ dashboard indicates that all the sealers use $\lit{tcp/30303}$.
Next, we simply extracted the autonomous system numbers (used to route the Internet traffic globally) based on their list of IP addresses.
After that, we used $\lit{nmap}$~\cite{noauthor_nmap:_nodate} to perform network OS fingerprinting on a list of IP addresses that we had already obtained.
Although results from network OS fingerprinting are not 100\% accurate, it was sufficient to map this information back to the OS names obtained from the dashboard because the cloud providers often use different kernel versions in their OS images.
Finally, we estimated the datacenters where sealers are deployed by looking at the proximities between their actual addresses and locations of the datacenters.
Figure~\ref{fig:core} summarizes the topological information we gathered from this inference.


\section{Experiments}
\label{sec:experiments}
In this section, we present the double spending results of the Cloning Attack in both Aura and Clique. 
For ethical reasons, we do not experiment our attack against PoA networks that are currently in production. Instead, we first present our experimental setup then detail the risk for an attacker to perform double spending in both Aura and Clique within our network.

\subsection{Testnet setup}\label{sec:test-setup}
To practically observe the chance of successful double spending using the approaches described in the previous sections we have created our own PoA blockchain networks, experimented the attacks and measured their success rate empirically.

Our testnet consists of 10 Ubuntu 18.04 Virtual Machines (VMs) on our OpenStack private cloud; each VM is provided with 1 virtual CPU core and 2 GB of memory.
These VMs are placed into two subnets, 5 VMs each; they are connected through 5 linux virtual routers and a physical Ethernet switch with dedicated VLAN.
An instance of either Parity-Ethereum 2.0.8 stable version with Aura or go-ethereum 1.8.21 stable version with Clique runs on each VM.

All of these instances are peering with each other to form the blockchain network.
While we have 10 Ethereum instances in total, our PoA blockchain employed only 9 unique private keys for sealers; the last instance instead uses the same key as the first one as explained in Section~\ref{sec:attacks} where 
one instance is seen as a clone of the other.
As of writing, neither Aura nor Clique incorporate a mechanism to prevent private key reusing.
One can simply copy a key and configuration files from one instance to another in order share the private key, other instances will simply accept the connection from the clone instance in the same way as the original.

In our experiments, the attacker is a Byzantine (or malicious) sealer with the intention to achieve double spending.
This attacker is provided the capability to transiently partition the network into two sealer groups: the attacker and the victim group.
We refer to the attacker group as the group of sealers whose blocks sealed during the network partition are intended to be adopted as a part of canonical chain, while we refer to the victim group as the group of sealers whose sealed blocks are intended to be discarded after the fork is resolved.

To grant the capability to partition the network, we allow our attacker to cut the network connectivity between two subnets using a firewall feature on the linux routers.
Note that the same result is achieved using a man-in-the-middle attack though ARP-spoofing in a local area network or with BGP-hijacking in other networks~\cite{EGJ18}.
The attacker is also provided the controllability over 2 Ethereum instances (2 VMs) that share the same private key used to seal the blocks.

The attacker aims to partition the network right before their turn to seal the block, where each sealer group must contain one VM that is under the control of the attacker.
To begin the attack, our attacker actively checks the owner of the current turn every 10\,ms in order to partition the network close to the right timing.
Right after the network partition, the attacker issues one transaction to each sealer group; these two conflicting transactions $\ms{TX_1}$ and $\ms{TX_2}$, for example Alice is giving all of her coins to Bob in $\ms{TX_1}$ and gives the same coins to Carol in $\ms{TX_2}$.

\begin{figure}[t]
	\centering
	\includegraphics[width=.5\textwidth]{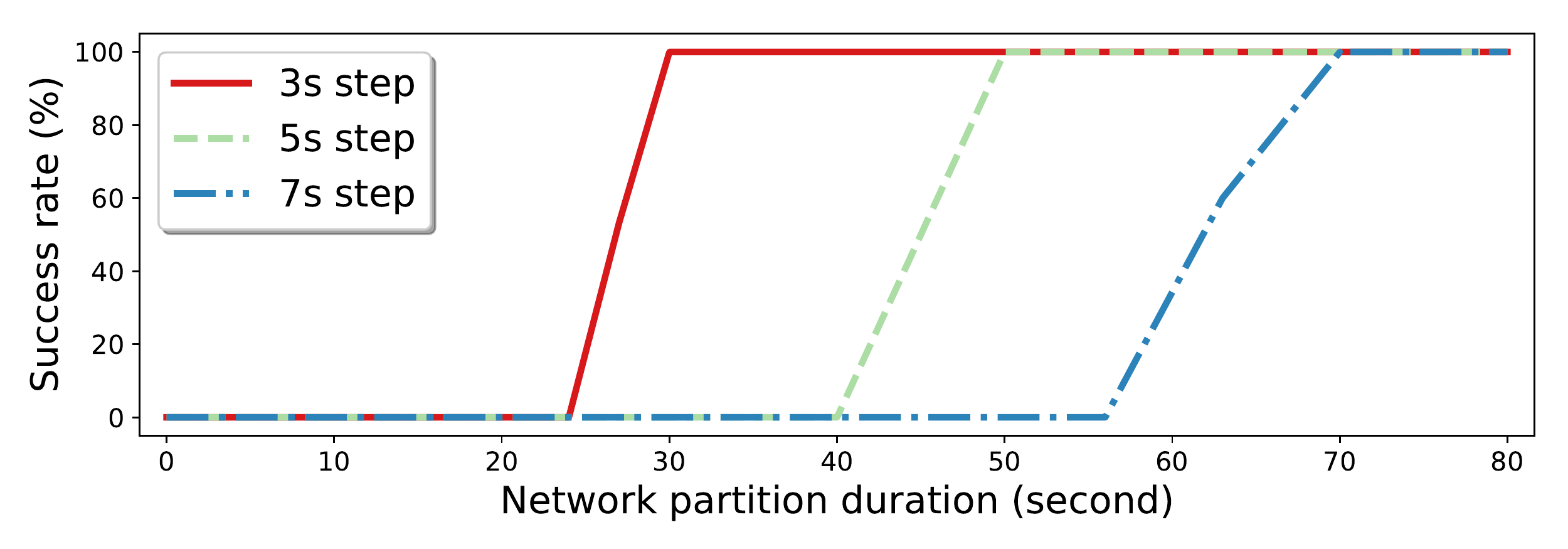}
	\caption{The success rate of double spending with the Cloning Attack in Aura}
	\label{fig:aura_success_rate}
\end{figure}

After issuing the transactions, the network partition is maintained during a period that depends on which PoA algorithm as explained below. 
When the fork is resolved at the end of the network partition, we look at the resulting branch of the fork which has been adopted as a canonical chain as well as the status of transactions.

A double spending is considered successful only if:
\begin{enumerate}
	\item A transaction issued to the victim group is committed before the end of the network partition;
	\item the blocks sealed by the attacker group during the fork have been adopted as a part of the canonical chain after the end of the partition; and
	\item the resulting canonical chain does not contain a transaction issued to the victim group.
\end{enumerate}

\subsection{Running the Cloning Attack against Aura}\label{sec:aura-experiment}

We experiment the Cloning Attack in Aura by varying the step duration and network partition duration.
We chose Step durations 3, 5, and 7 seconds in order to observe their impact of the minimum partition duration that makes the attack successful.
 
We maintain the network partition to match the step duration in use, such that for example a 24, 27, and 30 second partition duration corresponds respectively to the 8th, 9th and 10th step for a 3 second step duration, respectively.
We divide the sealers into two groups, such that apart from the two attacker instances, the placement of the reminder sealers is randomly but equally balanced between the two partitions. We do ensure, however, that both groups have an equal number of instances, which is 5, and each group contains one of two instances under the control of the attacker.
The values plotted for each combination of step duration and network partition duration are the averages over 30 runs.

\begin{figure}[t]
	\centering
	\includegraphics[width=.5\textwidth]{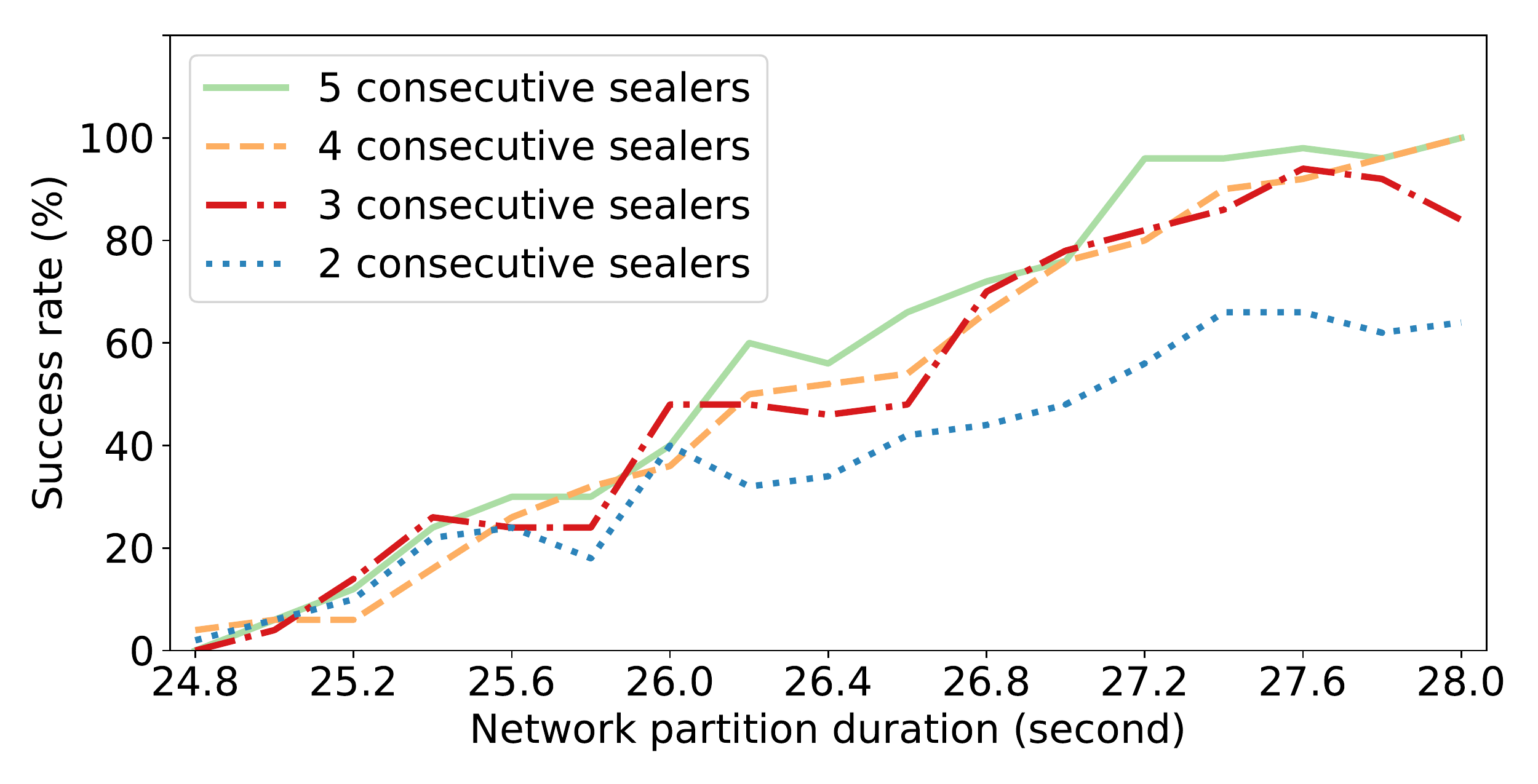}
	\caption{The success rate of the Cloning Attack double spending in Clique as the duration of the network partition increases and for different numbers of consecutive possible in-order sealers}
	\label{fig:clique_success_rate}
\end{figure}

\figurename~\ref{fig:aura_success_rate} presents the double spending success rate of 3, 5 and 7 second step durations.
In all three cases, the obtained results show a similar trend.
As expected, achieving a successful double spending is impossible in all step durations at the 8th step or any earlier step, namely 24, 40 and 56 seconds for 3, 5, and 7 second step durations, respectively.
Indeed, these attempts fail because any attack attempt among these runs could neither commit the transaction in the victim group nor force the block sealed by the attacker group to be adopted as a canonical chain when the network partition ends.

However, we can observe that as expected, the chances of successful double spending at the 9th step falls within the range between 50-60\%.
Even though both groups are provided enough time to seal 5 blocks in order to commit the transactions, the attacker still cannot force a particular branch of the fork to be adopted.
The variation at this point is due to the randomness of Ethereum instance placement during our experiment.

For all three step durations, at the 10th step and any step thereafter, the attack is always successful (100\% chances).
This is due to the attack technique in use that allows the attacker to force a branch of the fork to be adopted.
Overall, we can see that a longer step duration requires a longer period of network partition in order to achieve a successful double spending, which confirms our expectations.

\begin{figure}[t]
	\centering
	\includegraphics[clip=true,viewport=50 20 728 550,width=.55\textwidth]{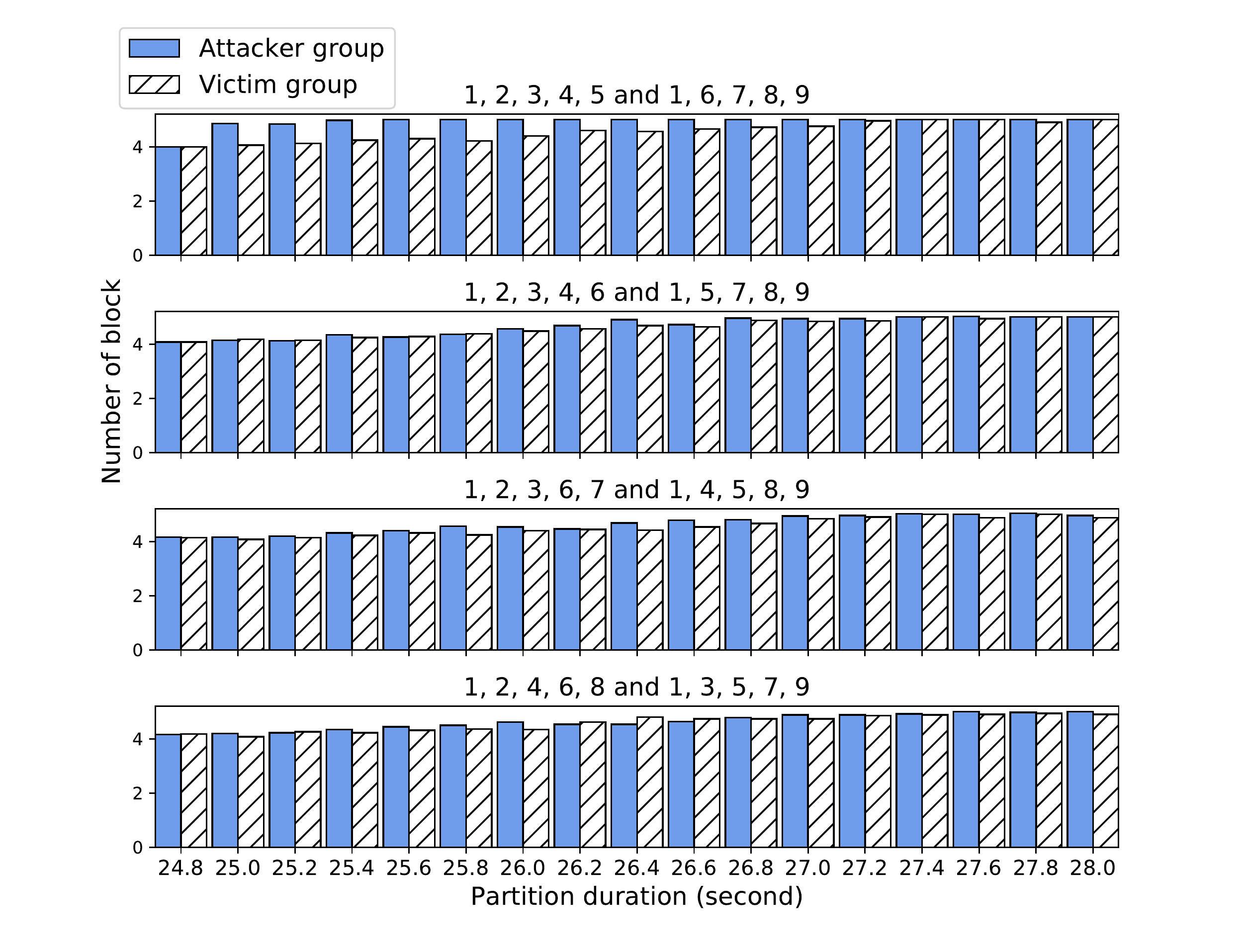}
	\caption{The average number of blocks created in Clique depending on the partition duration and the  sealer distribution across partitions}
	\label{fig:clique_block}
\end{figure}

\subsection{Running the Cloning Attack against Clique}\label{sec:clique-experiment}

We experiment the Cloning Attack on Clique while varying the partition duration and the way sealers are distributed between two partitions.

The variations in the sealer divisions are included in the experiments in order to capture the changes 
in weight of each branch as a result of the sealing sequences.
In particular, we experiment with the 4 sealer divisions presented below with different number of consecutive sealers:
\begin{itemize}
	\item 5 consecutive sealers: {1, 2, 3, 4, 5} in the attacker group and {1, 6, 7, 8, 9} in the victim group;
	\item 4 consecutive sealers: {1, 2, 3, 4, 6} in the attacker group and {1, 5, 7, 8, 9} in the victim group;
	\item 3 consecutive sealers: {1, 2, 3, 6, 7} in the attacker group and {1, 4, 5, 8, 9} in the victim group;
	\item 2 consecutive sealers: {1, 2, 4, 6, 8} in the attacker group and {1, 3, 5, 7, 9} in the victim group.
\end{itemize}

The partition duration is based on the block duration in use, which is fixed to 5 seconds in all our Clique experiments.
Since our testnet setup consists of 9 sealers in total, to commit a transaction during a partitioning, at least 5 blocks must be sealed in such a period.
In the best case where 5 sealers could seal 5 in-order blocks, the minimum duration required for the attack to succeed is equal to $5 \times 5 = 25$ seconds.
In other cases where at least 1 out of 5 blocks is sealed out-of-order, however, the required duration exceeds 25 seconds. 

Based on our knowledge of the time necessary for the algorithm to seal 5 blocks, we vary the duration from 24.8 to 28.0 seconds in an incremental step of 200 milliseconds. The range of duration allows to take into account the random delay of  out-of-order sealers as shown in Algorithm~\ref{algo:clique} and yet to capture the behavior of the system from the point where only 4 blocks can be sealed to the point where 5 blocks can be sealed. 

\begin{figure}[t]
	\centering
	\includegraphics[clip=true,viewport=50 20 728 550,width=.55\textwidth]{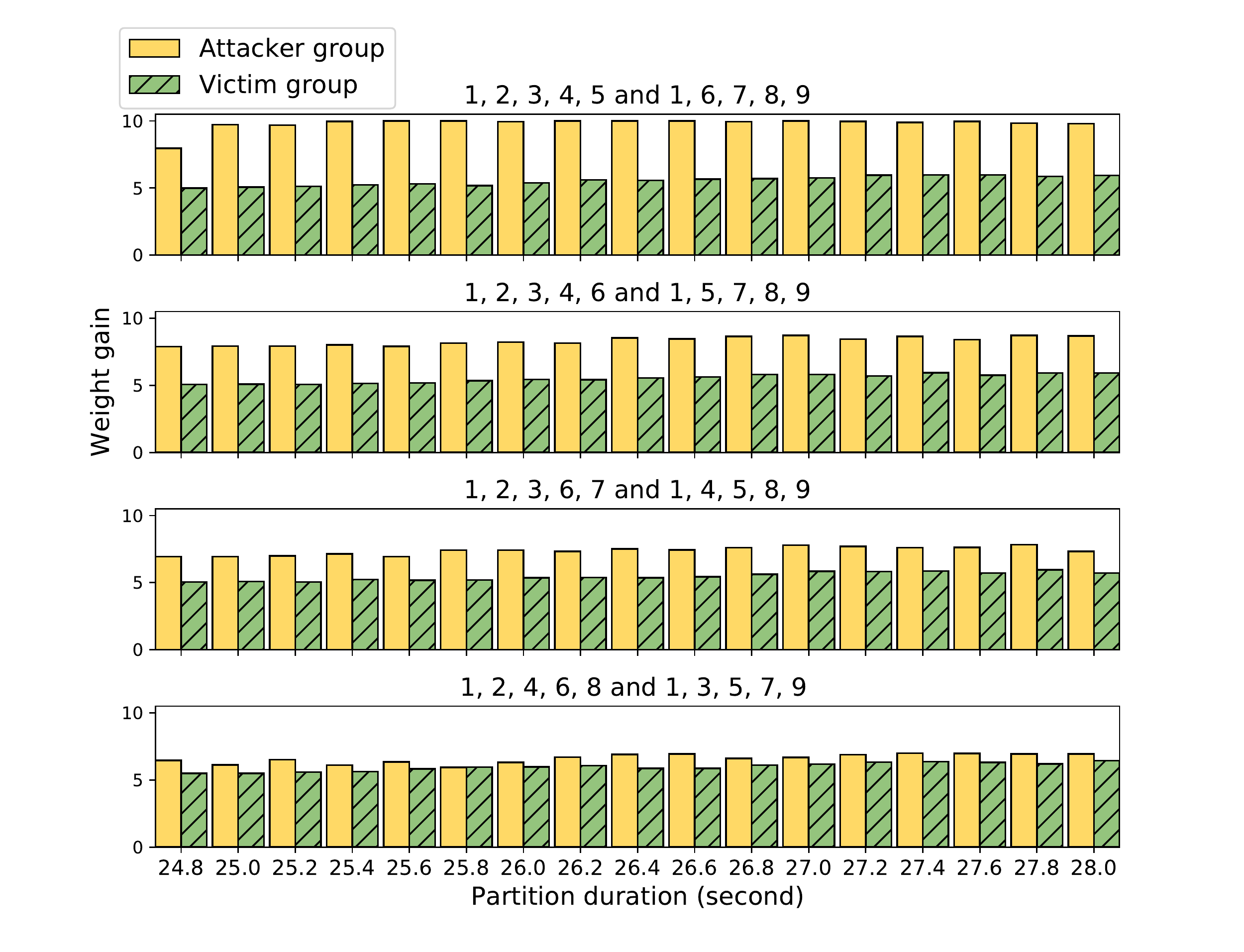}
	\caption{The average weight 
	gained 
	in Clique depending on the partition duration and the  sealer distribution across partitions}
	\label{fig:clique_diff}
\end{figure}

For each run we keep a record of whether the double spending was successful, which block was sealed by which sealer, the weight gained during the partition for each fork, and the number of blocks created during the network partition.
The  values averaged over 50 runs are depicted in the charts for each partition duration.


\figurename~\ref{fig:clique_success_rate} reveals the double spending success rate for the four aforementioned sealer divisions while \figurename~\ref{fig:clique_block} and \figurename~\ref{fig:clique_diff} show the number of sealed blocks and the weight gained during the network partition, respectively.
We observe that the success rate in \figurename~\ref{fig:clique_success_rate}, follows a similar trend for all of 4 grouping variations; the longer the partition duration, the higher the chance of successful double spending.

The shortest partition duration value in the chart, 24.8 second, gives the lowest success rate regardless of the sealer division.
This low success rate for short duration can be explained by the number of blocks sealed during the network partition.
Indeed, due to the limited partition duration, the victim group is rarely able to seal five blocks during the network partition as shown in \figurename~\ref{fig:clique_block}, thus a transaction issued by the victim group could not be committed and the attack fails.
When the partition takes longer, we can see that the victim group is able to seal five blocks.

When the partition duration is less or equal to 26 seconds, there is no noticeable difference between the four different sealer divisions.
In the case of two consecutive possible in-order sealers and when the partition duration is greater than 26 seconds, however, the success rate is lower than the other three divisions.
This phenomenon can be explained by the weight gained during the network partition as shown in \figurename~\ref{fig:clique_diff}.
In fact, in case of 2 consecutive possible in-order sealers, the difference in weight gained between attacker and victim  branches becomes relatively low; this gap narrows with the increase in the partition duration.

\section{Analysis and countermeasures} \label{sec:analysis}

We begin this section by comparing the vulnerabilities of Aura and Clique to the Cloning Attack resulting from our experiments in Section~\ref{sec:experiments}.
Next, we analyse further the Aura algorithm and discuss its implication to the blockchain safety and liveness.
Finally, we present potential countermeasures against the Cloning Attack. 

\subsection{Comparison between Aura and  Clique}\label{sec:comparison}

In this section, we explain why the Cloning Attack against Aura can always be successful whereas
the Cloning Attack against Clique is much faster but not always successful.

As detailed in Section~\ref{sec:consensus}, one of the main differences between Aura and Clique resides in the predictability of the sequence of sealers. 
In fact, in Aura the sequence is strictly enforced whereas in Clique this sequence may change depending on the difference between a random number and the network communication delay.
This slight algorithmic difference has however significant consequences on consensus algorithms resilience to double spending attacks using our proposed Cloning Attacks.

On the one hand and as we have demonstrated in Section~\ref{sec:experiments},  due to its strict enforcement of sealing order, Aura is  vulnerable to the Cloning Attack in case of network partition. 
Performing the Cloning Attack against Aura, the attacker does not need to know anything about the identity of the sealers nor does it need to know their order. 
Thus, a malicious sealer only needs to partition the overlay network using classical network attacks such as BGP hijacking to succeed in double spending with a 100\% chance of success. 

On the other hand, double spending without topology information on Clique is possible, but 
the attack against Clique is about twice as fast as against Aura when the topology is known.
Indeed, as we have presented in Section~\ref{sec:experiments}, the knowledge of potential next in-order sealer greatly influences the chance of double spending.
When the attacker is capable of isolating the next $\lfloor n/2 \rfloor +1$ sealers, it is able to perform the double spending attack with 100\% success rate. 
On the opposite, the knowledge of only the next two in-order sealers only guarantees a success rate of  60\% maximum.

Interestingly, when considering the attacks against both Aura and Clique without the knowledge of the topology, it appears that attacking Clique can be even slower than attacking Aura.
The reason is that in the worst case scenario where all in-order sealers are on the victim side, the attacker will have to obtain a branch that is twice as large as the victim branch before it can double spend. Growing this branch would take more time than executing the Cloning Attack on Aura.
But overall, even without knowledge of the topology both Aura and Clique consensus algorithms are vulnerable to a malicious sealer aiming at double spending.



\subsection{Requirements to make Aura safe and live}\label{sec:liveness-safety}

Our attack violates the safety of Aura, in that it leads the system to an undesirable state, where coins can be stolen.\footnote{Safety (resp. liveness) is often referred to as the property of a system to guarantee that  "a bad thing will never happen" (resp. "a good thing eventually happens")~\cite{lamport1977,alpern1987recognizing}.}
As we explain here, this problem can be mitigated by simply increasing the number of sealers $|V|$ necessary to decide a block.  
%
Determining $|V|$ to ensure safety may however be insufficient, as it does not ensure that the 
system makes progress, which is a liveness problem. So we also explain how to choose $|V|$
to ensure that Aura is both safe and live.
Changing $|V|$ can be easily achieved by modifying the boolean condition under which a block is decided at line~\ref{line:aura-is-decided} of Algorithm~\ref{algo:aura}.

\begin{figure}[t]
	\centering
	\includegraphics[clip=true, viewport=35 20 1500 890,width=.52\textwidth]{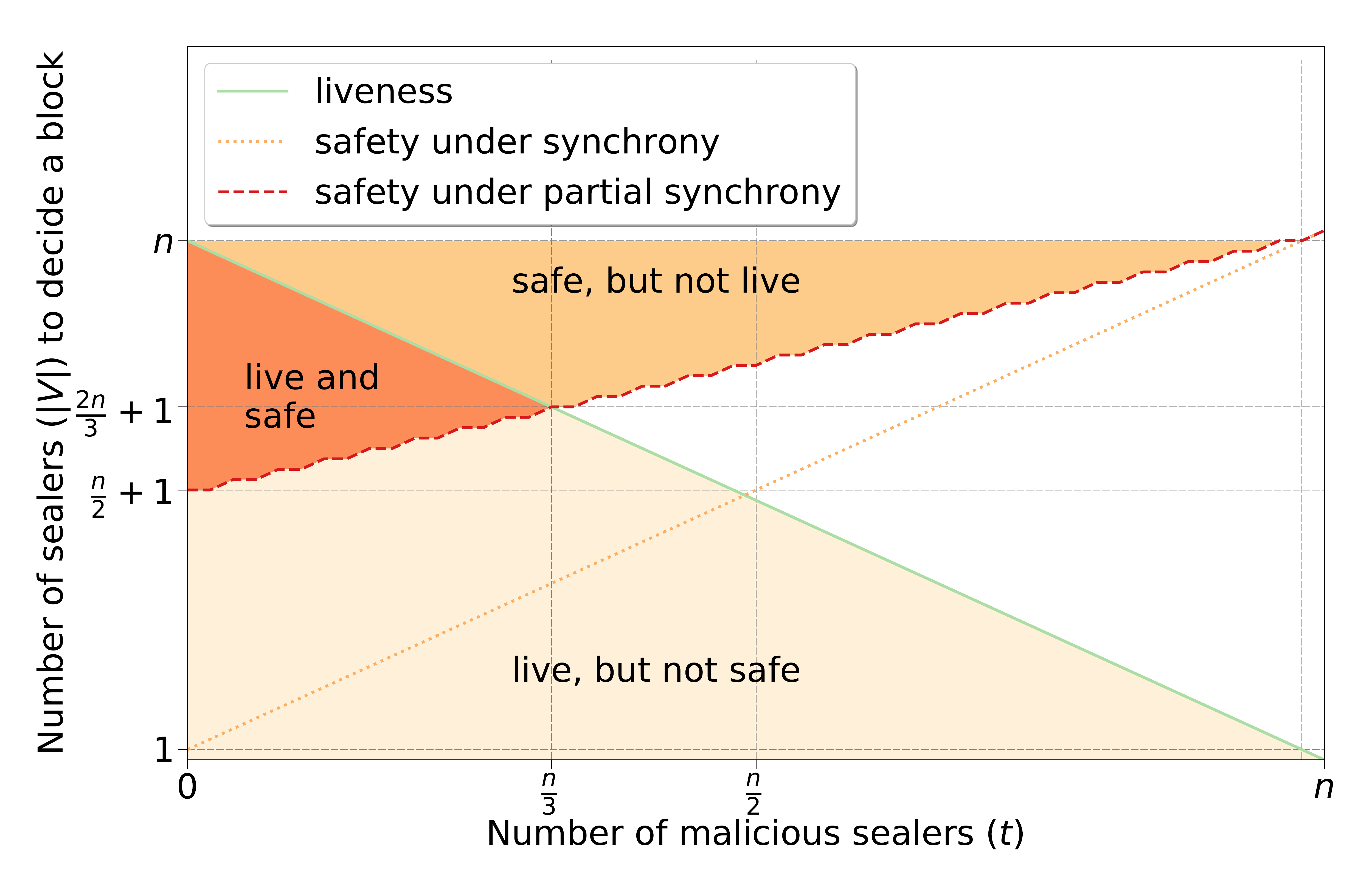}
	\caption{Required number of sealers $|V|$ to decide a block depending on the number of malicious sealers $t$: Aura is safe only when $|V| >  \frac{n+t}{2}$, Aura is live only when $|V| < n - t$, so Aura is both safe and live only when $\frac{n + t}{2} < |V| <  n-t$.}	
	\label{fig:Byzantine_blocks_relationship}
\end{figure}

Figure~\ref{fig:Byzantine_blocks_relationship} depicts the relation between the desired fault tolerance 
of the system and the number of sealers  $|V|$  necessary to decide a block to ensure  Aura's safety.
This analysis is helpful as it allows us depending on the targeted fault tolerance, to decide the minimal number of sealers that are necessary to decide a block in a synchronous or partially synchronous network so as to ensure that Aura is safe and live.
In particular, we consider two distinct cases, whether the network communication is synchronous or partially synchronous. The asynchronous case is ignored here as consensus would be impossible in all interesting cases where $t>0$~\cite{FLP85}.\\

\noindent
{\bf Synchronous network.} In the synchronous case, Aura's safety is guaranteed when 
$|V| > t$, which is represented by the area above the orange dotted line $|V|=t+1$ in Figure~\ref{fig:Byzantine_blocks_relationship}. In fact, to ensure that there exists a block sealed by an honest sealer, one has to wait for the number of blocks sealed by distinct sealers to be greater than the total number of malicious sealers, which requires the number of sealers $|V|$ to exceed the number $t$ of malicious sealers.
  \begin{itemize}
	\item {\bf Case $t = 0$:} When all the sealers are honest, $|V| = 1$ is sufficient to consider finality of a block, i.e., the transactions within a block can be considered committed instantly because one may safely assume that a block will be delivered to all the sealers in the known upper bounded time.
	\item {\bf Case $0 < t < n$:} When there is at least one malicious sealer, one must ensure that at least one honest sealer has recently appended a block to the blockchain before considering finality of its previous blocks because a malicious sealer may violate the protocol by introducing a malformed block or appending a block out-of-order.
	\item {\bf Case $t = n$:} In this case, we can see that $|V|$ should be strictly greater than $n$, which is impossible by definition. This illustrates that it is impossible for Aura to work 
	when $t=n$.
  \end{itemize}
  To conclude the highest fault tolerance $t$ that Aura can tolerate in a synchronous system is $t=n-1$ by requiring blocks from all the $|V| = n$ sealers to decide a block.\\

\noindent
{\bf Partially synchronous network.}
Requiring $\lfloor \frac{n}{2} \rfloor + 1$ sealers to decide one block is insufficient  
to tolerate unpredictable message delays between two partitions of honest sealers when $t >0$.
As an example, recall that Aura aims at tolerating up to a minority $\lceil \frac{n}{2}\rceil -1$ of malicious sealers, however, 
if a majority $\lfloor \frac{n}{2} \rfloor + 1$ of sealers are sufficient to decide a block, then a group of only 2 honest sealers helped with the $t=\lfloor \frac{n}{2} \rfloor + 1$ malicious sealers would be sufficient to decide one block. This does not prevent the other honest sealers in another partition from deciding a conflicting block with the help of the clones of the $t$ malicious sealers. 
This is the reason why the dashed red line on Figure~\ref{fig:Byzantine_blocks_relationship} indicates that $|V| \geq \lfloor \frac{n+t}{2} \rfloor +1$
is necessary to guarantee that a majority of honest sealers sign a block so that no other conflicting blocks can be decided. More precisely, here are the 3 interesting cases to consider:

\begin{itemize}
\item {\bf Case $t = 0$.} When all the sealers are honest, the algorithm only needs at least a majority of sealers, 
$|V| \geq \lfloor\frac{n}{2}\rfloor + 1$, 
before considering that a block is decided.
\item {\bf Case $0 < t < \frac{n}{3}$.} 
To ensure that a majority of the honest sealers seal a block for one block to be decided, 
we need $|V| \geq \lfloor \frac{n+t}{2} \rfloor +1$ sealers to 
seal blocks. As there are $n-t$ honest sealers, a majority of them contains $\lfloor \frac{n-t}{2} \rfloor +1$ sealers. As there are $t$ malicious sealers,
we need strictly more than half $\frac{n-t}{2}$ of the honest sealers and the $t$ malicious ones to seal blocks, which leads to 
$|V| \geq \lfloor \frac{n+t}{2} \rfloor +1$. 
The two upper triangles of Figure~\ref{fig:Byzantine_blocks_relationship} depict these conditions, under which Aura is safe.
\item {\bf Case $t \geq \frac{n}{3}$.} 
Interestingly, when $t \geq \frac{n}{3}$, it is impossible to guarantee that at least $|V| \geq \lfloor \frac{n+t}{2} \rfloor +1$ will seal blocks. In fact, this would imply that strictly more than $\frac{2n}{3}$ sealers seal a block. However, as $t \geq \frac{n}{3}$ this would also mean that at least one malicious sealer seals a block, which cannot be guaranteed as malicious sealers may choose, by definition, to not follow the protocol. The upper-right triangle of Figure~\ref{fig:Byzantine_blocks_relationship} depicts the conditions under which Aura is safe because $|V| \geq \lfloor \frac{n+t}{2} \rfloor +1$ but not live because $t \geq \frac{n}{3}$.
\end{itemize}
Even with only one malicious sealer ($t=1$), the current implementation of Aura cannot guarantee  safety.
The reason is that Aura claims that a majority of sealers is enough as long as $t\leq \frac{n}{2}$.
However, one can find a counter-example where $|V| = \lfloor\frac{n}{2}\rfloor + 1$
that falls in the unsafe area shown in Figure~\ref{fig:Byzantine_blocks_relationship}.

\begin{figure}[t]
	\centering
	\includegraphics[width=.5\textwidth]{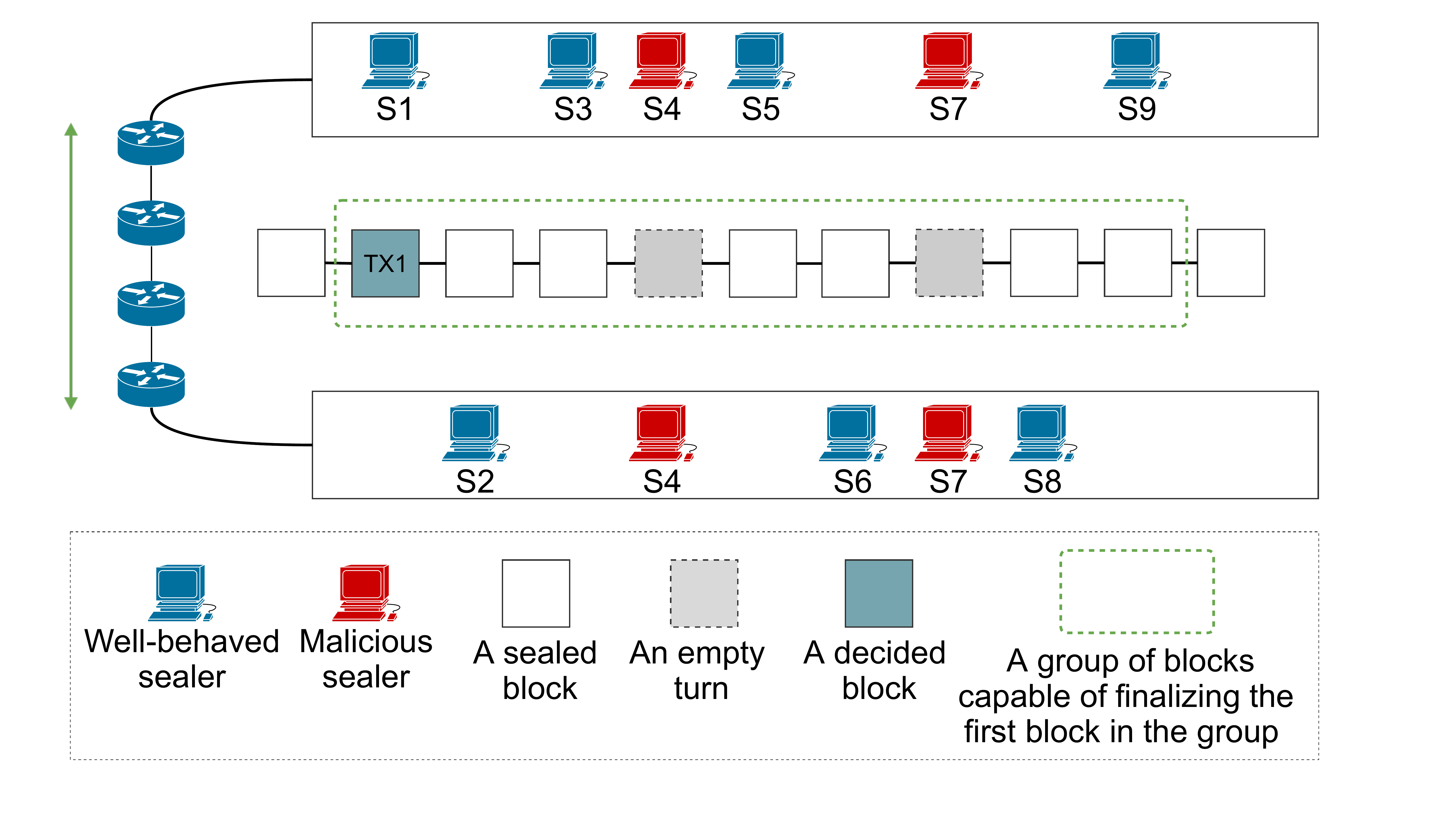}
	\caption{An Aura execution using $|V| > \frac{2n}{3}$ and $t<\frac{n}{3}$ ($n=9$, $|V|=7$ and $t=2$) where transaction $\ms{TX1}$ is committed}
	\label{fig:aura_2-3}
\end{figure}

\subsection{Simple safe countermeasure}\label{sec:countermeasure}

As explained in the introduction and for ethical reasons, we disclosed early a simple, yet naive, countermeasure to make Aura safe. This countermeasure has been implemented in the xDai blockchain as acknowledged in their white paper~\cite{BAF19}.
This first countermeasure was not as precise as of today and required $|V|>\frac{2n}{3}$ and $t<\frac{n}{3}$ (our technical report is not cited here for the sake of anonymity) to ensure safety but ignoring liveness guarantees.  Interestingly, the same conditions of this countermeasure are required by IBFT~\cite{jpmorganchase_quorum} and recent work suggested that IBFT is not live either~\cite{saltini2019correctness}.
%

\begin{figure}[t]
	\centering
	\includegraphics[width=.5\textwidth]{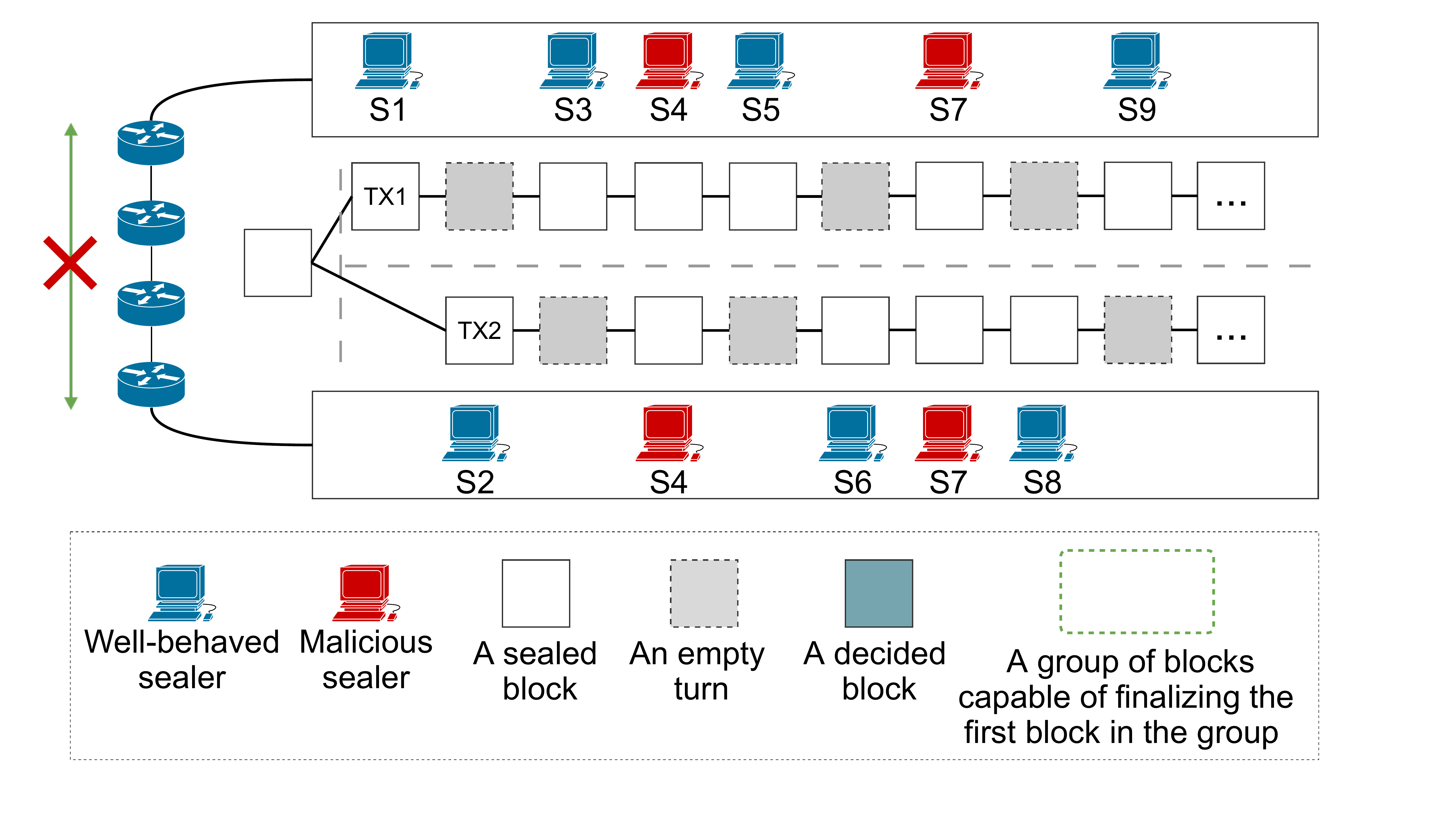}
	\caption{
	An Aura execution using $|V| > \frac{2n}{3}$ and $t<\frac{n}{3}$ ($n=9$, $|V|=7$ and $t=2$) where  neither transaction $\ms{TX1}$ nor $\ms{TX2}$ are committed}
	\label{fig:aura_2-3_partition}
\end{figure}

As an example, \figurename{s}~\ref{fig:aura_2-3} and~\ref{fig:aura_2-3_partition} demonstrate how a modified version of Aura where  $n=9$, $|V| = 7$, $t=2$ where S4 and S7 and malicious.
These two sealers are allowed to be silent or even seal the blocks in all groups whenever a network partition occurs. 
%
As indicated with a green dash frame in \figurename~\ref{fig:aura_2-3}, this version of Aura can commit $\ms{TX_1}$ even though S4 and S7 do not contribute any block to the chain in their turns.
The S4 turn can be left empty and S5 may simply continue sealing a block after that; the same goes for S8 after the end of S7 turn.
In this case, the other 7 honest sealers alone are sufficient to decide a block.
%
As indicated in the execution of \figurename~\ref{fig:aura_2-3_partition}, however, 
 it is impossible to have at least 7 sealers on both sides at the same time.
Therefore, it is impossible to commit the transactions issued to both partitions concurrently, even though Aura allows their sealers to continue sealing more blocks during the network partition.

\subsection{Safe and live countermeasures}\label{sec:countermeasure2}

On Figure~\ref{fig:Byzantine_blocks_relationship}, it is easy to identify that the simple countermeasure presented above, which requires $|V|>\frac{2n}{3}$ and $t<\frac{n}{3}$, does not ensure liveness. This is because it actually tolerates conditions in the upper-right triangle that do not ensure liveness.

To remedy this issue, it is sufficient to ensure that $\frac{n + t}{2} < |V| <  n-t$. Note that this implicitly guarantees that $t < \frac{n}{3}$ anyway. As a result, under these conditions Aura remains safe and live.

Finally, a radically different countermeasure that also offers safety and liveness is to
use a deterministic consensus algorithm that is partially synchronous in that it tolerates arbitrary delays. PBFT~\cite{CL99} is one example as it relies on a leader but is not designed to scale outside a small network. DBFT~\cite{CGLR18} is a leaderless deterministic partially synchronous consensus algorithm that was especially designed to scale to blockchain systems.
%
In addition, DBFT is time optimal and resilience optimal.
It has been shown that DBFT is resilient to double spending attacks, as it is not possible for a blockchain  building upon it, like the Red Belly Blockchain~\cite{CNG18}, to fork.


\section{Conclusion}\label{sec:conclusion}

To cope with the drawbacks of proof-of-work, Byzantine fault tolerance has been introduced in mainstream blockchains in the form of proof-of-authority where sufficient sealers $|V|$ among $n$ must seal a block despite a minority $t$ of them being malicious. 
The Cloning Attack allows malicious participants to double spend in Ethereum instances using Clique and Aura consensus protocols.
Our findings inspired by the theory of Byzantine fault tolerance define precisely the necessary and sufficient condition $\frac{n + t}{2} < |V| <  n-t$ under which PoA is safe and live.
%
The Cloning Attack consists of duplicating private keys for cloning and 
require that the network messages are delayed. We explained 
how to gather the necessary topological information to identify how to attack the underlying network of some of these blockchains.

\section*{Acknowledgements}
We wish to thank Kirill Pimenov, Head of Security at Parity Technologies, for going through the technicalities of the attack of the clones and confirming the theoretical vulnerability of the $\lit{parity}$ software and to Martin Holst Swende from Ethereum for confirming the possibility of the attack on $\lit{geth}$.
This research is in part supported under Australian Research Council Discovery Projects funding scheme (project number 180104030) entitled ``Taipan: A Blockchain with Democratic Consensus and Validated Contracts''. Vincent Gramoli is a Future Fellow of the Australian Research Council.

\bibliographystyle{plain}
\bibliography{reference}

\end{document}